\documentclass[journal]{new-aiaa}
\usepackage[utf8]{inputenc}

\usepackage{xcolor}
\usepackage[version=4]{mhchem}
\usepackage{longtable,tabularx}
\title{Lie Group Structure of Six-Degree-of-Freedom Rigid-Body Uncertainty Representation and Propagation}

\author{Zhuochen Wang\footnote{PhD student, School of Mathematics; email: zwang3005@gatech.edu.} and Panagiotis Tsiotras\footnote{Professor, Guggenheim School of Aerospace Engineering; email: tsiotras@gatech.edu.}}
\affil{Georgia Institute of Technology, Atlanta, GA, 30332}

\usepackage{amsthm}
\usepackage{bbm}
\usepackage{siunitx}
\usepackage{algorithm}
\usepackage{algpseudocode}
\usepackage{enumitem}
\usepackage{quiver}
\usepackage{xparse}
\usepackage{mathrsfs}
\usepackage{cancel}
\usepackage{bm}
\usepackage{xparse}
\usepackage{makecell}
\usepackage[percent]{overpic}
\usepackage{tabularx}
\usepackage{booktabs}
\usepackage{float} 

\newcommand{\B}{\mathcal{B}}
\newcommand{\rr}{\ensuremath{\mathbb{R}}}

\newcommand{\norm}[1]{{\left\lVert{#1}\right\rVert}}

\newcommand{\II}{\mathbb{I}} 
\newcommand{\defeq}{:=}

\newcommand{\so}{\ensuremath{\mathfrak{so}}}

\newcommand{\deri}[2][]{{\mathrm{T}_{#1} \mathord{#2}}} 
\newcommand{\tg}[2][]{\mathrm{T}_{#1} \mathord{#2}} 

\newcommand{\ha}{{^\wedge}}
\newcommand{\ve}{{^\vee}}

\newcommand{\defemph}[1]{{\emph{#1}}}
\NewDocumentCommand\lsd{D<>{\Ad^{-1}}O{1}}{\ltimes_{#1_{#2}}}
\NewDocumentCommand\rsd{D<>{\Ad}O{1}}{\rtimes_{#1_{#2}}}
\NewDocumentCommand{\vframe}{ m o o }{%
  {#1}%
  \IfValueT{#2}{%
    \IfBlankF{#2}{_{\scriptscriptstyle #2}}%
  }%
  \IfValueT{#3}{%
    \IfBlankF{#3}{^{\scriptscriptstyle #3}}%
  }%
}

\newcommand{\pureq}{\mathbb{H}_p}
\newcommand{\unitq}{\mathbb{H}_u}

\NewDocumentCommand{\dqmult}{O{1}}{\odot_{#1}}

\NewDocumentCommand\ldqmul{O{1}}{\ltimes_{#1}}
\NewDocumentCommand\rdqmul{O{1}}{\rtimes_{#1}}

\newcommand{\TSEql}{\mathrm{TSE^q}(3)^{\ell}}

\newcommand{\eAd}{\bm{\Ad}}

\newcommand{\SOq}{\ensuremath{\mathrm{SO^q}(3)}}
\newcommand{\SEq}{\ensuremath{\mathrm{SE^q}(3)}}
\newcommand{\JJ}{\mathbb{J}}

\DeclareMathOperator{\tr}{tr}

\DeclareMathOperator{\Rot}{Rot}
\DeclareMathOperator{\atan2}{atan2}
\DeclareMathOperator{\sinc}{sinc}
\DeclareMathOperator{\Ad}{Ad}
\DeclareMathOperator{\ad}{ad}
\DeclareMathOperator{\SO}{SO}
\DeclareMathOperator{\supp}{supp}
\DeclareMathOperator{\blkdiag}{blkdiag}

\DeclareMathOperator{\TSO}{TSO}

\DeclareMathOperator{\SE}{SE}
\DeclareMathOperator{\TSE}{TSE}

\DeclareMathOperator{\expm}{expm}

\DeclareMathOperator{\ANEES}{ANEES}

\let\svlim\lim\def\lim{\svlim\limits}

\usepackage{scalerel,stackengine}
\stackMath
\newcommand*\wh[1]{%
\savestack{\tmpbox}{\stretchto{%
  \scaleto{%
    \scalerel*[\widthof{\ensuremath{#1}}]{\kern-.6pt\bigwedge\kern-.6pt}%
    {\rule[-\textheight/2]{1ex}{\textheight}}
  }{\textheight}%
}{0.5ex}}%
\stackon[1pt]{#1}{\tmpbox}%
}
\let\epsilon\varepsilon
\let\phi\varphi

\RequirePackage{hyperref} 
    \hypersetup{
        colorlinks=true,
        linkcolor=blue,
        citecolor=blue
    }

\RequirePackage[capitalize,nameinlink]{cleveref}[0.19]

\crefname{section}{section}{sections}
\crefname{subsection}{subsection}{subsections}
\crefname{appendix}{Appendix}{Appendices}
\Crefname{section}{Section}{Sections}
\Crefname{subsection}{Subsection}{Subsections}
\Crefname{appendix}{Appendix}{Appendices}
\Crefname{figure}{Figure}{Figures}

\crefformat{equation}{\textup{#2(#1)#3}}
\crefrangeformat{equation}{\textup{#3(#1)#4--#5(#2)#6}}
\crefmultiformat{equation}{\textup{#2(#1)#3}}{ and \textup{#2(#1)#3}}
{, \textup{#2(#1)#3}}{, and \textup{#2(#1)#3}}
\crefrangemultiformat{equation}{\textup{#3(#1)#4--#5(#2)#6}}%
{ and \textup{#3(#1)#4--#5(#2)#6}}{, \textup{#3(#1)#4--#5(#2)#6}}{, and \textup{#3(#1)#4--#5(#2)#6}}

\Crefformat{equation}{#2Equation~\textup{(#1)}#3}
\Crefrangeformat{equation}{Equations~\textup{#3(#1)#4--#5(#2)#6}}
\Crefmultiformat{equation}{Equations~\textup{#2(#1)#3}}{ and \textup{#2(#1)#3}}
{, \textup{#2(#1)#3}}{, and \textup{#2(#1)#3}}
\Crefrangemultiformat{equation}{Equations~\textup{#3(#1)#4--#5(#2)#6}}%
{ and \textup{#3(#1)#4--#5(#2)#6}}{, \textup{#3(#1)#4--#5(#2)#6}}{, and \textup{#3(#1)#4--#5(#2)#6}}

\crefdefaultlabelformat{#2\textup{#1}#3}

\theoremstyle{plain}
\newtheorem{theorem}{Theorem}
\newtheorem{lemma}{Lemma}

\newtheorem{prop}{Proposition}

\theoremstyle{definition}

\theoremstyle{remark}

\usepackage{etoolbox}
\AtEndEnvironment{vb}{\null\hfill$\diamond$}%
\AtEndEnvironment{intermezzo}{\null\hfill$\diamond$}%
\renewcommand{\d}{\ensuremath{\mathrm{d}}}

\begin{document}

\maketitle

\begin{abstract}

For six-degree-of-freedom rigid-body motion, the choice of Lie-group structure affects both how well the group-theoretic mean and covariance represent the state distribution and how accurately these moments are propagated.
A quaternion-based, left-trivialized semidirect-product formulation of the velocity phase space is developed for the actuated body-frame dynamics, with closed-form group operations and Jacobians that avoid matrix-group embeddings.
We derive the controlled stochastic dynamics and establish that the direct- and semidirect-product formulations describe the same physical process under the specified noise model.
Since the semidirect exponential coordinates account for frame changes between body twists at different poses, they capture pose--twist coupling more naturally.
A lunar powered-descent experiment assesses low-order moment representations of uncertainty and propagation performance under the proposed group structure using two distinct algorithms.
Compared with direct-product formulations and a linearized baseline, the proposed structure improves the representation of the banana-shaped body-velocity distribution using group-theoretic moments, yields more consistent propagated moments, and produces distribution approximations that agree more closely with the Monte Carlo reference.
In this experiment, changing the group structure accounts for most of the observed improvement, highlighting the practical value of the proposed semidirect-product formulation.

\end{abstract}

\section{Introduction}

Distributional control seeks to shape the evolution of a system's state distribution. 
In aerospace applications, covariance steering has been used to control the terminal state dispersion during powered descent \cite{ridderhof2018uncertainty}. 
For nonlinear six-degree-of-freedom (6DOF) rigid-body dynamics, however, an initially Gaussian distribution can become highly distorted in physical coordinates and thus it may be poorly described by just its mean and covariance in these coordinates. 
Monte Carlo simulation and higher-order moment methods can capture this distortion, but can be computationally demanding. 
An alternative is to seek coordinates in which the evolving distribution remains well described by its first two moments. This requires both a useful uncertainty representation and accurate moment propagation.

As a first step toward distributional control for actuated rigid-body system in many aerospace engineering applications, we consider uncertainty propagation for 6DOF actuated rigid-body motion. 
A standard approach is the prediction step of the multiplicative extended Kalman filter (MEKF) \cite{lefferts1982kalman}, which propagates covariance by linearizing about the estimated mean trajectory. 
This linearized approximation also underlies nonlinear covariance steering \cite{Ridderhof2019,yu2025covariance} and stochastic 6DOF spacecraft trajectory optimization \cite{Zhang2023Stochastic6DOFRendezvous}. 
However, its accuracy can deteriorate as the distribution is distorted over time by nonlinear rigid-body dynamics on the underlying Lie group structure. 
This observation motivates Lie-theoretic alternatives, including sigma-point methods \cite{Brossard2020UKFM,Mangiacapra2022UKFControlTSE3} and intrinsic mean--covariance equations from a second-moment closure of the Fokker--Planck equation evolution \cite{ye2024uncertaintya}. 
%

The velocity and momentum phase spaces are the tangent and cotangent bundles of the configuration space (e.g., attitude and position), respectively \cite{marsden1999introduction}. 
For 6DOF motion, we focus on the velocity phase space $\TSE(3)$ of the configuration group $\SE(3)$.
In deterministic treatments, the group structure of this phase space is often left implicit, since the established equations of motion suffice for propagation. 
In stochastic settings, however, the choice of group structure affects how uncertainty is represented and propagated.

There is no consistent choice of phase-space Lie-group structure across the guidance, navigation, and control (GNC) literature. 
Direct-product formulations appear in Lie-group filtering on $\SO(3)\times\rr^9$ \cite{Bourmaud2015CDLGEKF} and $\SO(3)\times\rr^3$ \cite{Cesic2017ExtendedInformationFilter}, pose--velocity filtering on $\TSE(3)$ \cite{Mangiacapra2022UKFControlTSE3}, and attitude--momentum uncertainty propagation \cite{ye2024uncertaintya}. 
By contrast, semidirect-product structures are used in phase-space uncertainty propagation \cite{Jayaraman2023LieTheoreticUncertainty} and equivariant filter design \cite{Fornasier2022EqFInputBiases,Fornasier2025EquivariantSymmetriesINS}. 
The latter structures explicitly capture the coupling between configuration and velocity or momentum variables and are consistent with the natural semidirect-product structure of phase spaces \cite{marsden1999introduction}.

The choice of the group structure determines the coordinates in which uncertainty is represented. 
For example, coupled attitude and position uncertainty can produce banana-shaped distributions that are approximately Gaussian in the exponential coordinates of an appropriate Lie group \cite{LongRSS2012Banana,barrau2020imu}. 
This suggests choosing a semidirect-product structure on $\TSE(3)$ whose exponential coordinates similarly capture the coupling among pose, angular velocity, and linear velocity.

Related equivariant formulations use right-trivialized matrix Lie groups tailored to inertial navigation \cite{Fornasier2022EqFInputBiases,Fornasier2025EquivariantSymmetriesINS}. 
In contrast, we adopt a left-trivialized formulation aligned with body-frame dynamics and actuation noise, particularly relevant to strongly actuated maneuvers such as powered descent. 
In addition, a quaternion representation of attitude avoids the computational overhead of a cumbersome matrix-group embedding \cite{dantam2021}.

To the best of our knowledge, there is no prior quaternion-based treatment of \emph{left}-trivialized $\TSE(3)$, which we denote as $\TSEql$. 
We examine how this structure affects uncertainty representation and propagation under the same physical stochastic dynamics. 
We develop the formulation within an abstract Lie-theoretic framework, with the following contributions.

\begin{enumerate}
    \item 
    We identify and formalize the natural left-trivialized Lie-group structure of the velocity phase space over general Lie groups, thereby clarifying a geometric point that is treated inconsistently in the existing literature.
    
    \item 
    For 6DOF rigid-body motion, we develop the quaternion-based 
    $\TSEql$ and derive efficient closed-form group operations, adjoint operators, exponential and logarithmic maps, and their Jacobians.
    
    \item 
    We derive the controlled \textit{stochastic} rigid-body dynamics on $\TSEql$ and establish equivalence of the physical process under direct- and semidirect-product representations for the specified noise model.
    
    \item 
    We demonstrate, using a lunar powered-descent experiment, improved concentrated Gaussian representation of body-velocity uncertainty and more consistent and accurate uncertainty propagation, with the proposed group structure accounting for the dominant gains.
\end{enumerate}
\section{Mathematical Preliminaries}

\subsection{Lie Groups and Their Tangent Bundles}

We recall the constructions needed below and refer the interested reader to \cite{dummit2004abstract,kirillov2008introduction,holm2011geometric} for additional background and details.
Let $G$ be an $n$-dimensional Lie group with identity element $e$ and Lie algebra $\mathfrak g=\tg[e]{G}$, where $\mathfrak g$ is endowed with a Lie bracket $[\cdot,\cdot]$ induced by the group structure of $G$. 
We use $\tg[e]{G}$ when emphasizing
the additive group structure of $\mathfrak g$ or the corresponding Lie algebra structure with zero bracket. 
For $x\in\rr^n$, a chosen basis $\{E_i\}_{i=1}^n$ of $\mathfrak g$ defines the hat map $x\ha=x_iE_i \in \mathfrak g$ and the vee map $(x\ha)\ve = x$, where repeated indices are implicitly summed throughout the paper. 
Lie-algebra elements are identified with their coordinate vectors when this is unambiguous from the context.

\begin{figure}[h]
    \centering
    \includegraphics[width=0.65\linewidth]{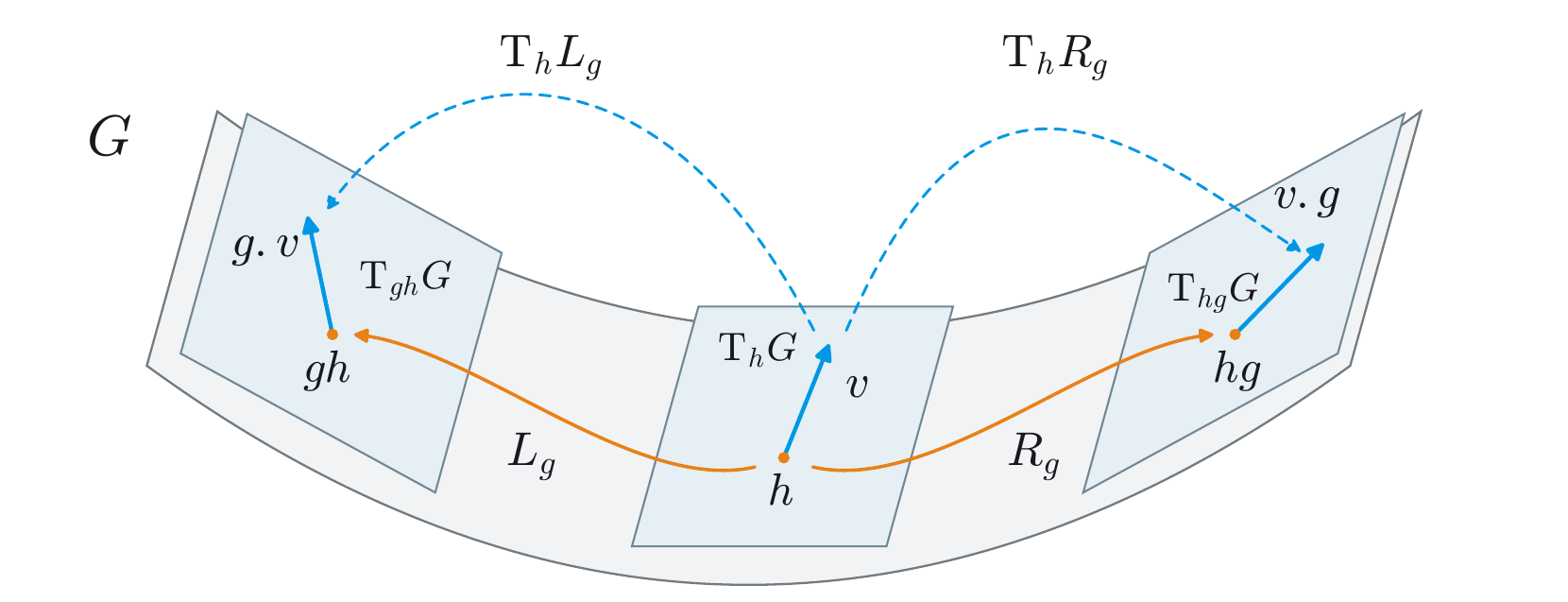}
    \caption{Left and right translations of tangent vectors on $G$.}
    \label{fig:translations}
\end{figure}

The left and right multiplication on $G$, defined by $L_g(h)=gh$ and $R_g(h)=hg$, induce \emph{left and right translation actions} on the tangent vectors: $g.v\defeq\deri[h]{L_g}(v)$ and $v.g\defeq\deri[h]{R_g}(v)$ for $v\in\tg[h]{G}$
(see Figure~\ref{fig:translations}). The notation $\tg[g]{G}$ denotes the tangent space of $G$ at $g$, whereas $\tg[g]{f}$ denotes the derivative of a smooth map $f$ at $g$.

Throughout, the ``." denotes these translation actions. 
The conjugation operation on $G$ is defined as  $C_g(h)=ghg^{-1}$ for $g,h \in G$.
Differentiating the conjugation yields the \emph{group adjoint action} $\Ad_g\defeq\deri[e]{C_g}$, and differentiating $\Ad$ yields the \emph{algebra adjoint} $\ad_\xi(\eta)=[\xi,\eta]$ for $\xi,\eta \in \mathfrak g$.
We use square brackets for the corresponding coordinate matrices so we can write $[\Ad_g]\xi\ve=(\Ad_g(\xi))\ve$ and $[\ad_\xi]\eta\ve=[\xi,\eta]\ve$.

The tangent bundle $\tg[]{G}=\{(g,v):g\in G,\ v\in\tg[g]{G}\}$ inherits the group multiplication from $G$ and the product rule:
\begin{align}
    (g_1,v_1)*(g_2,v_2)
    =(g_1g_2,\,g_1.v_2+v_1.g_2).
\end{align}
The
\emph{right and left trivialized $\tg[]{G}$} are respectively the semidirect-product groups from the Lie-group isomorphisms
\begin{subequations}
\begin{align}
    \textrm{right:}\quad \tg{G} \ni (g,v)&\mapsto(g,v.g^{-1})
    \in G\rsd<\Ad>[]\tg[e]{G},\\
    \textrm{left:}\quad \tg{G} \ni (g,v)&\mapsto(g,g^{-1}.v)
    \in G\lsd<\Ad^{-1}>[]\tg[e]{G}.
\end{align}
\end{subequations}
The corresponding semidirect-product laws of the two trivialized groups are
\begin{subequations}
\begin{align}
    \textrm{right:}\quad
    (g_1,r_1)*(g_2,r_2)
    &=(g_1g_2,\,r_1+\Ad_{g_1}(r_2)),
    \label{eq:rsdp}\\
    (g,r)^{-1}&=(g^{-1},-\Ad_{g^{-1}}(r)),
    \label{eq:rsdp_inv}\\
    \textrm{left:}\quad
    (g_1,r_1)*(g_2,r_2)
    &=(g_1g_2,\,\Ad_{g_2}^{-1}(r_1)+r_2),
    \label{eq:lsdp}\\
    (g,r)^{-1}&=(g^{-1},-\Ad_g(r)).
    \label{eq:lsdp_inv}
\end{align}
\end{subequations}
For example, for rigid-body motion on $G=\SE(3)$, right and left trivialization of the tangent vector $v\in\tg[g]{G}$ give the \textit{spatial twist} $\vframe{v}[S]=(v.g^{-1})\ve$ and the \textit{body twist} $\vframe{v}[B]=(g^{-1}.v)\ve$, respectively, with $\vframe{v}[S]=[\Ad_g]\vframe{v}[B]$. 
We use the same notation for operations on $G$ and its trivialized tangent bundles, with the group determined by the arguments.

\begin{prop}\label{prop:lr_tri_ad}
The group adjoint matrices in the right and left trivialized $\tg{G}$ are
\begin{align}
    \textrm{right:}\quad
    [\Ad_{(g,r)}]
    &=\begin{bmatrix}
        [\Ad_g]&0\\
        [\ad_r][\Ad_g]&[\Ad_g]
    \end{bmatrix},
    \label{eq:rtri_Ad_mat}\\
    \textrm{left:}\quad
    [\Ad_{(g,r)}]
    &=\begin{bmatrix}
        [\Ad_g]&0\\
        [\Ad_g][\ad_r]&[\Ad_g]
    \end{bmatrix},
    \label{eq:ltri_Ad_mat}
\end{align}
and both trivializations have the same algebra adjoint
\begin{align}
    [\ad_{(\xi,v)}]
    =\begin{bmatrix}
        [\ad_\xi]&0\\
        [\ad_v]&[\ad_\xi]
    \end{bmatrix}.
    \label{eq:tan_tri_ad}
\end{align}
The corresponding group adjoint actions of the right and left trivialized groups are
\begin{align}
    \textrm{right:}\quad
    \Ad_{(g,r)}(\xi,v)
    &=(\Ad_g(\xi),\,\Ad_g(v)+\ad_r(\Ad_g(\xi))),
    \label{eq:rtri_Ad}\\
    \textrm{left:}\quad
    \Ad_{(g,r)}(\xi,v)
    &=(\Ad_g(\xi),\,\Ad_g(v+\ad_r(\xi))).
    \label{eq:ltri_Ad}
\end{align}
\end{prop}
These expressions follow by differentiating the conjugation maps in the two trivialized groups. The same algebra adjoint shared by the two trivialized groups induces a shared Lie algebra, which we denote
by $\mathfrak g\rsd<\ad>[]\tg[e]{G}$ or $\mathfrak g\lsd<\ad>[]\tg[e]{G}$ according to the trivialization. 

\begin{lemma}\label{lem:left_tri_tan}
For $H=G\lsd<\Ad^{-1}>[]\tg[e]{G}$, left trivialization of $(\dot g,\dot r)\in\tg[(g,r)]{H}$ gives
\begin{align}
    (g,r)^{-1}.(\dot g,\dot r)
    =(g^{-1}.\dot g,\,\dot r+\ad_{g^{-1}.\dot g}(r)).
    \label{eq:left_action_ltan}
\end{align}
If $g^{-1}.\dot g=r$ holds, then we have
\begin{align}
    (g,r)^{-1}.(\dot g,\dot r)=(r,\dot r).
    \label{eq:left_action_ltan_reduce}
\end{align}
\end{lemma}
\begin{proof}
Differentiating Eq.~\eqref{eq:lsdp} gives $(g,r).(\xi,v)=(g.\xi,v-\ad_\xi(r))$. Solving for $(\xi,v)$ gives Eq.~\eqref{eq:left_action_ltan}. The second identity follows from the property $\ad_r(r)=0$.
\end{proof}

\subsection{Exponential Coordinates and Jacobians}

Let $\exp_G:\mathfrak g\to G$ denote the Lie-group exponential map~\cite{kirillov2008introduction}. 
We use the corresponding \defemph{coordinate exponential map} 
$\exp(x)\defeq\exp_G(x\ha)$ and its local inverse $\log:O\to D$, where $\exp:\rr^n \supset D \to O\subset G$ is restricted to a neighborhood $D$ on which it is a diffeomorphism. 
The \defemph{exponential coordinates of $g$ centered at $g_0$} are $x=\log(g_0^{-1}g)$, and we can recover $g=g_0\exp(x)$.

The left and right Jacobians $J_\ell$ and $J_r$ express the differential of $\exp$ in right- and left-trivialized $\mathfrak g$ coordinates, respectively: for $y\in\rr^n$,
\begin{align}
    J_\ell(x)y
    =\bigl((\deri[x]{\exp}(y)).\exp(-x)\bigr)\ve,
     \qquad J_r(x)y
    =\bigl(\exp(-x).(\deri[x]{\exp}(y))\bigr)\ve.
    \label{eq:Jr}
\end{align}
Writing $X=[\ad_{x\ha}]$ and denoting the matrix exponential by $\expm$, their matrix formulas are \cite{Iserles2000LieGroupMethods}
\begin{align}
    J_\ell(x)=\int_0^1\expm(sX)\,\d s
    =\sum_{k=0}^{\infty}\frac{X^k}{(k+1)!}, \qquad J_r(x)=J_\ell(-x).
    \label{eq:mat_Jl}
\end{align}
They are invertible on $D$:
\begin{align}
    J_\ell^{-1}(x)
    =-\frac12 X+\sum_{k=0}^{\infty}
      \frac{B_{2k}}{(2k)!}X^{2k}, \qquad J_r^{-1}(x)=J_\ell^{-1}(-x),
    \label{eq:mat_Jl_inv}
\end{align}
where $B_k$ are the Bernoulli numbers. 
Since the two trivialized groups share the same algebra adjoint by Proposition~\ref{prop:lr_tri_ad}, 
their Jacobians have the same expression.

\begin{prop}[{\cite[Theorem 3.1]{moskowitz2003exponential}}]
\label{prop:tan_tri_exp}
The coordinate exponential and logarithmic maps in the right and left trivialized groups are
\begin{align*}
    \textrm{right:}\quad
    \exp(\eta,\nu)&=(\exp(\eta),J_\ell(\eta)\nu), \qquad
    \log(g,\xi)=(\log(g),J_\ell^{-1}(\log(g))\xi),\\
    \textrm{left:}\quad
    \exp(\eta,\nu)&=(\exp(\eta),J_r(\eta)\nu), \qquad
    \log(g,\xi)=(\log(g),J_r^{-1}(\log(g))\xi).
\end{align*}
\end{prop}

These Jacobians also express group dynamics in exponential coordinates.
\begin{prop}[{\cite[Corollary 9]{MuntheKaas1999HighOrderRKManifolds}}]
\label{prop:rkmk}
If $g(t)$ solves the left-trivialized ODE
\begin{align}
    (g^{-1}.\dot g)\ve=f(g,t),
    \qquad g(t_0)=g_0,
    \label{eq:group_ode}
\end{align}
then locally $g(t)=g_0\exp(x(t))$, where
\begin{align*}
    \dot x=J_r^{-1}(x)f(g_0\exp(x),t),
    \qquad x(t_0)=0.
\end{align*}
\end{prop}
Runge--Kutta--Munthe-Kaas (RKMK) methods \cite{MuntheKaas1999HighOrderRKManifolds} apply a Runge--Kutta scheme to this equation in exponential coordinates and map the update back through $\exp$.

\subsection{Rigid-Body Representations and Reconstruction}
\label{sec:rigid_body_reconstruction}

For $\SO(3)$, we use the standard basis $E_i=e_i\ha$ in $\so(3)$, where $x\ha y=x\times y$ for $x,y \in \rr^3$. 
Then, for $R \in \SO(3)$, we have $[\Ad_R]=R$ 
and $[\ad_{x\ha}]=x\ha$. 
For rotations with angle less than $\pi$, the principal logarithm in rotation-vector coordinates takes values in $D=\B_\pi^3$, the open 3-ball of radius $\pi$ centered at the origin. 
Because the group adjoint action of $\SO(3)$ is its usual action on $\rr^3$, the rigid-motion group $\SE(3)=\SO(3)\rsd<\Ad>[]\rr^3$ is identified with the right-trivialized $\TSO(3)$. 
Standard formulas for $\SO(3)$ and $\SE(3)$ are collected in Appendix~\ref{app:formulas}.

We take $R$ to map body coordinates to spatial coordinates, and $r=\vframe{r}[S]$ to be the spatial position. The \defemph{reconstruction equation} $g^{-1}.\dot g=\xi$ relates the pose rate to the body twist $\xi=(\vframe{\Omega}[B],\vframe{v}[B])$ \cite{marsden1999introduction}. For $g=(R,r)$, it reads
\begin{align}
    \vframe{\Omega}[B]\ha=R^\top\dot R,
    \qquad
    \vframe{v}[B]=R^\top\vframe{\dot r}[S][S],
    \label{eq:recon_eq_se3}
\end{align}
where the notation $\vframe{\dot r}[A][F]$ denotes differentiation of the position vector in frame $F$, with the result expressed in basis $A$. 

For computation, we represent rotations by unit quaternions $q=(q_s,q_v) \in \unitq$ \cite{markley2004attitude}. 
We suppress the hat and vee notations for pure quaternions $\pureq$ by identifying $\pureq$ with $\rr^3$, denote Hamilton quaternion multiplication by $\otimes$ or juxtaposition, and write $q^* = q^{-1}$ for the unit quaternion conjugate.
This way, for any $\eta \in \rr^3$, $\eta\ha$ always unambiguously denotes the algebra adjoint matrix of $\so(3)$ in this paper. 
The covering map $\Rot:\unitq\to\SO(3)$ that converts a unit quaternion into a rotation matrix identifies $q$ and $-q$ as the same rotation. 
Let $\SOq$ denote this quaternion representation of $\SO(3)$, with $\{\pm q\}$ treated as the same group element. 
We preserve the standard $\so(3)$ coordinates: a vector $\eta$ in the standard $\so(3)$ coordinates corresponds to the pure-quaternion coordinates $\eta/2$. 
Treating $\vframe{\Omega}[B]$ as a pure quaternion, we have
\begin{align}
    \dot q&=q.\left( \frac{1}{2}\vframe{\Omega}[B]\right)=\frac{1}{2}q\otimes\vframe{\Omega}[B]
    ,
     \qquad
    \vframe{\Omega}[B]= 2(q^*. \dot q)=2q^*\otimes\dot q ,
    \qquad
    \Ad_q(r)=qrq^*=\Rot(q)r.
    \label{eq:Adq}
\end{align}
The coordinate $\exp$ and $\log$ maps are
\begin{align}
    \exp(\eta)
    &=\left(
        \cos(\theta/2),
        \ \sinc(\theta/2)\eta/2
    \right),
    && \theta=\norm{\eta},\\
    \log(q)&=
    2\atan2(\norm{q_v},q_s)\frac{q_v}{\norm{q_v}}, && q_s > 0,
\end{align}
where $\sinc(z)=\sin(z)/z$ with its continuous value at zero. For the logarithm, we choose the representative $q$ with $q_s>0$ to keep the output in $D$ and set $\log(1)=0$ by continuity.

The corresponding representation $\SEq=\SOq\rsd<\Ad>[]\rr^3$ has operations
\begin{align}
    (q_1,r_1)*(q_2,r_2)&=(q_1q_2,r_1+\Ad_{q_1}(r_2)),
    \qquad
    (q,r)^{-1}=(q^*,-\Ad_{q^*}(r)).
\end{align}
Its adjoint is obtained by substituting $R=\Rot(q)$ above, and its exponential and logarithmic maps are
\begin{align}
    \exp(\eta,\rho)
    =(\exp(\eta),J_\ell(\eta)\rho),
    \qquad
    \log(q,r)
    &=(\log(q),J_\ell^{-1}(\log(q))r).
    \label{eq:se3_log}
\end{align}
Quaternion representations preserve the standard Lie-algebra coordinates, so the algebra adjoint and hence the left and right Jacobians and their inverses remain unchanged from those of $\SO(3)$ and $\SE(3)$. 
The reconstruction equation is 
\begin{align}
    (q,\vframe{r}[S])^{-1}.
    (\dot q,\vframe{\dot r}[S][S])
    =(\vframe{\Omega}[B]/2,\vframe{v}[B]).
    \label{eq:recon_eq_se3q}
\end{align}

\subsection{Distribution on a Unimodular Lie Group}
\label{sec:dist}

A Lie group $G$ is \defemph{unimodular} if $|\det[\Ad_g]|=1$ for every $g\in G$. 
Its left and right Haar measures coincide, giving a bi-invariant measure denoted by $\d g$ \cite[Secs.~10.5.4 and 12.1.1]{chirikjian2012stochastic}. 
This bi-invariant Haar measure provides a natural reference for defining probability densities, preserves volume under left and right translations, and allows us to directly apply the uncertainty propagation results of \cite{ye2024uncertaintya}.
The block matrices in Proposition~\ref{prop:lr_tri_ad} give $\det[\Ad_{(g,r)}]=(\det[\Ad_g])^2$. 
Thus, if $G$ is unimodular, its right- and left-trivialized tangent bundles are also unimodular. 
Therefore, the rotation and rigid-motion groups in this paper are unimodular.

For a density $p$ with respect to Haar measure, a \defemph{group-theoretic mean} $\mu$ and its \defemph{group-theoretic covariance} $\Sigma$ satisfy
\begin{align}
    &\int_G\log(\mu^{-1}g)p(g)\,\d g=0,
    \label{eq:group_mean}\\
    \Sigma&=\int_G\log(\mu^{-1}g)
    \log(\mu^{-1}g)^\top p(g)\,\d g.
    \label{eq:group_cov}
\end{align}
We call the mean-centered exponential coordinates $\log(\mu^{-1}g)$ \defemph{log-error coordinates}. 
We assume that the mean exists and is locally unique,
that $\log$ is defined on $\mu^{-1}\supp p$, and that
the log errors have finite second moment.
These moments depend on the group operations and logarithm. 
Changing the group structure on the same manifold can therefore change both the group-theoretic mean and covariance. 
These moments also generally differ from distance-based moments \cite{khan2025means,PennecArsigny2013ExponentialBarycenters}.

A \defemph{concentrated Gaussian distribution on $G$} is  parameterized by $\mu$ and $\Sigma$ via $g=\mu\exp(x)$, where $x\sim\mathcal N(0,\Sigma)$.
We use it in the concentrated regime, where the probability
of $x$ lying outside the domain $D$ is negligible.
The support condition above is then understood approximately, and $\mu$ and $\Sigma$ represent the group-theoretic mean and covariance under this approximation.
Later, we assess how well the group-theoretic mean and covariance describe a propagated distribution by comparing it with the concentrated Gaussian approximation constructed from these moments.
The accuracy of this approximation depends on how nearly
Gaussian the propagated distribution is in log-error coordinates.

\subsection{Stochastic Processes on Lie Groups}
We consider group-valued stochastic processes governed by Stratonovich SDEs with smooth coefficients and Wiener drivers, and refer readers to \cite{chirikjian2012stochastic,hsu2002stochastic,CatuognoRuffino2003StochasticExponential,ye2024uncertaintyb} for background. 
Stratonovich calculus provides a natural formulation of physical stochastic dynamics on Lie groups because its chain rule allows us to express the dynamics in the Lie algebra by left trivialization.
We then convert the resulting SDE to It\^{o} form and
recover the same group-valued stochastic process through a
limit of left-endpoint exponential products.
We elaborate on this construction below.

Let a $G$-valued process $g_t$ satisfy the Stratonovich SDE
\begin{align}
    \d g_t
    =V_0(g_t,t)\,\d t
    + V_i(g_t,t)\circ\d W_t^i,
    \label{eq:group_stratonovich}
\end{align}
where $V_0,\ldots,V_m$ are smooth vector fields on $G$ and $W_t$ is a standard $m$-dimensional Wiener process.
We assume a prescribed initial condition $g_0$ and that the
solution exists on the time interval of interest.
The ``$\circ$'' denotes Stratonovich integration.
When manipulating the Stratonovich differential geometrically, we write
$\circ\d g_t$, while keeping the conventional notation
$\d g_t$ on the left-hand side of Eq.~\eqref{eq:group_stratonovich}.
For any smooth map $\Phi$, the Stratonovich differential
satisfies the usual chain rule $\circ\d\Phi(g_t)=\d\Phi_{g_t}(\circ\d g_t)$.
Thus $\circ\d g_t$ transforms as a tangent vector
\cite{Norris1992}, allowing it to be left-trivialized to the Lie algebra.

The \defemph{left stochastic logarithm} of $g_t$ is the $\mathfrak g$-valued process
\begin{align}
    M_t=\int_0^t g_s^{-1}.(\circ\d g_s),
    \qquad M_0=0.
\end{align}
The stochastic logarithm accumulates left-trivialized increments and generally differs from the pointwise coordinate logarithm $\log(g_0^{-1}g_t)$.
Left trivializing Eq.~\eqref{eq:group_stratonovich} therefore gives
\begin{align}
    \d(M_t\ve)
    =f^s(g_t,t)\,\d t+F^s(g_t,t)\circ\d W_t,
\end{align}
where, with $n=\dim G$,
$f^s(g,t)\in\rr^n$ and $F^s(g,t)\in\rr^{n\times m}$ are
the left-trivialized coefficients in $\mathfrak g$ coordinates.

Conversely, the \defemph{left stochastic exponential} of $M_t$
is the solution $h_t$ of $\d h_t=h_t.(\circ\d M_t)$ with $h_0=e$.
The original process is recovered as $g_t=g_0h_t$.
Equivalently, the \defemph{McKean--Gangolli injection}
maps increments of $M_t$ to group increments $\exp((M_{t_{k+1}}-M_{t_k})\ve)$ and accumulates them by successive right multiplication, recovering $g_t$
in the limit of vanishing time steps
\cite{chirikjian2012stochastic,ye2024uncertaintyb}.

Since $M_t\ve$ is an $\rr^n$-valued process, ordinary Stratonovich--It\^{o} conversion gives the equivalent It\^{o} representation
\begin{align}
    \d(M_t\ve)=f(g_t,t)\,\d t+F(g_t,t)\,\d W_t,
\end{align}
where
\begin{align}
    f=f^s+\frac12 E_i^r(F_{kj}^s)F_{ij}^s e_k,
    \qquad F=F^s,
    \label{eq:Ito_correction}
\end{align}
and, for smooth $\phi:G\to\rr$, the right Lie derivative  along the basis element $E_i$ of $\mathfrak g$ is
\begin{align}
    E_i^r(\phi)(g)
    \defeq\left.\frac{\d}{\d\tau}\right|_{\tau=0}
    \phi\bigl(g\exp(\tau e_i)\bigr),
    \qquad e_i=E_i\ve.
\end{align}
Because $F^s$ depends on $g_t$, the correction accounts for
its variation along the noise directions on $G$.

Using these It\^{o} coefficients, the McKean--Gangolli construction is formally written as
\begin{align}
    g_{t+\d t}=g_t\exp\bigl(
        f(g_t,t)\,\d t+F(g_t,t)\,\d W_t\bigr).
\end{align}
This denotes the continuous-time limit of successive exponential updates with coefficients evaluated at the left endpoints. We use the shorthand
\begin{align}
    (g_t^{-1}\d g_t)\ve
    =f(g_t,t)\,\d t+F(g_t,t)\,\d W_t.
\end{align}
Here $(g_t^{-1}\d g_t)\ve$ denotes the It\^{o} differential $\d(M_t\ve)$ through the McKean--Gangolli construction. 
This way, the final equation represents the same group-valued stochastic process as Eq.~\eqref{eq:group_stratonovich}.
\section{Rigid-Body Motion on the 6DOF Velocity Phase Space}

The \defemph{velocity phase space} of 6DOF rigid-body motion is the tangent bundle $\TSE(3)$. We shall derive its left trivialization because rotational dynamics and body-fixed actuation are naturally expressed in body coordinates.

\subsection{Mixed Trivialization and Group Structure}

The rigid-body representation in Section~\ref{sec:rigid_body_reconstruction} allows the pose and body twist to be treated through right trivializing $\tg{\SO(3)}$ into $\SE(3)$ and then left trivializing $\TSE(3)$. 
These mixed-directional, successive trivializations should yield the natural group structure of spatial pose and body twist that we seek. 
We first describe this construction for a general Lie group $G$. Let $H\defeq G\rsd<\Ad>[]\tg[e]{G}$ be the right trivialized $\tg[]{G}$. 
Left-trivializing $\tg[]{H}$ gives $H\lsd<\eAd^{-1}>[]\tg[(e,0)]{H}$, where $\eAd$ denotes the adjoint of $H$. 
Writing its elements as $(g,r,\Omega,v) \in H\lsd<\eAd^{-1}>[]\tg[(e,0)]{H}$, we first need the group multiplication law to understand its structure:
\begin{equation}
\begin{aligned}
    (g_1,r_1,\Omega_1,v_1)*(g_2,r_2,\Omega_2,v_2)=\bigl(g_1g_2,\ r_1+\Ad_{g_1}(r_2),
        \Ad_{g_2^{-1}}(\Omega_1)+\Omega_2,
        \Ad_{g_2^{-1}}\bigl(v_1+\ad_{\Omega_1}(r_2)\bigr)+v_2\bigr).
    \label{eq:double_tri}
\end{aligned}
\end{equation}
To obtain this group law,
we use $(g,r)^{-1}=(g^{-1},-\Ad_{g^{-1}}(r))$ in $H$ and apply Proposition~\ref{prop:lr_tri_ad}:
\begin{align}
    \Ad_{(g,r)^{-1}}(\Omega,v)
    &=\bigl(\Ad_{g^{-1}}(\Omega),
        \Ad_{g^{-1}}(v)
        +\ad_{-\Ad_{g^{-1}}(r)}(\Ad_{g^{-1}}(\Omega))\bigr)\\
    &=\bigl(\Ad_{g^{-1}}(\Omega),
        \Ad_{g^{-1}}(v+\ad_\Omega(r))\bigr).
\end{align}
The second equality follows from properties 
of the Lie bracket. 
Substituting this adjoint into Eq.~\eqref{eq:lsdp} gives Eq.~\eqref{eq:double_tri}.

Specializing the construction to the left-trivialized $\TSE(3)$, we obtain a group whose elements have entries as attitude, spatial position, body angular velocity, and body linear velocity. 
For computation, we use the quaternion-based $\TSEql\defeq\SEq\lsd<\eAd^{-1}>[]\tg[(1,0)]{\SEq}$. Equation~\eqref{eq:double_tri} gives
\begin{align}
    (q_1,r_1,\Omega_1,v_1)*(q_2,r_2,\Omega_2,v_2)&=\bigl(q_1q_2,\ r_1+\Ad_{q_1}(r_2),
        \Ad_{q_2^*}(\Omega_1)+\Omega_2,
        \Ad_{q_2^*}(v_1+\Omega_1\times r_2)+v_2\bigr),
    \label{eq:tse3q}\\
    (q,r,\Omega,v)^{-1}
    &=\bigl(q^*,\ -\Ad_{q^*}(r),\ -\Ad_q(\Omega),
        -\Ad_q(v)+\Ad_q(\Omega)\times r\bigr).
    \label{eq:tse3q_inv}
\end{align}
Notice that the group law naturally encodes the Transport Theorem, accounting for the frame dependence of body twist.

Throughout the remainder of the paper, we denote the following direct- and semidirect-product groups by
\begin{align}
    G_{rd}\defeq\SEq\times\rr^6,
    \qquad G_{r\ell}\defeq\TSEql.
    \label{eq:phase_space_groups}
\end{align}
The subscripts identify the attitude--position and pose--twist products, respectively. 
Both groups have the same underlying manifold, with elements $h=(g,\xi)$, but different group operations. In $G_{rd}$, twists compose by addition; in $G_{r\ell}$, the adjoint action accounts for their dependence on the relative pose.

For $G_{r\ell}$, using the standard coordinate basis of $\mathfrak{g}_{r\ell}\cong\rr^{12}$, Proposition~\ref{prop:lr_tri_ad} yields
\begin{align}
    [\Ad_{(q,r,\Omega,v)}]
    =\begin{bmatrix}
        R&0&0&0\\
        r\ha R&R&0&0\\
        R\Omega\ha&0&R&0\\
        r\ha R\Omega\ha+Rv\ha&R\Omega\ha&r\ha R&R
    \end{bmatrix},
    \qquad [\ad_{(\eta,\rho,\omega,\nu)}]
    =\begin{bmatrix}
        \eta\ha&0&0&0\\
        \rho\ha&\eta\ha&0&0\\
        \omega\ha&0&\eta\ha&0\\
        \nu\ha&\omega\ha&\rho\ha&\eta\ha
    \end{bmatrix},
    \label{eq:tse3q_ad}
\end{align}
where $R=\Rot(q)$ and each block is $3\times3$. The coordinate exponential and logarithmic maps follow from Proposition~\ref{prop:tan_tri_exp}:
\begin{align}
    \exp(\eta,\rho,\omega,\nu)
    &=\left(
        \exp(\eta,\rho),
        J_r(\eta,\rho)[\omega,\nu]^\top
    \right),
    \label{eq:tse3q_exp}\\
    \log(q,r,\Omega,v)
    &=\begin{bmatrix}
        \log(q,r)\\
        J_r^{-1}(\log(q,r))\begin{bmatrix}\Omega\\v\end{bmatrix}
    \end{bmatrix}.
    \label{eq:tse3q_log}
\end{align}
The $\SEq$ maps and Jacobians on the right-hand sides are given in the preliminaries and Appendix~\ref{app:formulas}.

\begin{figure}[htbp]
    \centering
    \includegraphics[width=0.55\linewidth]{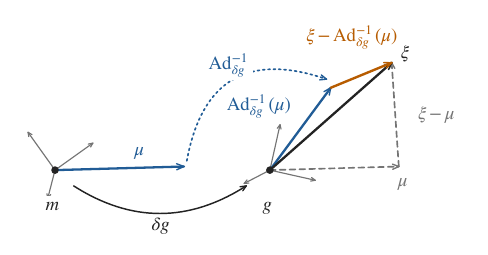}
    \caption{Euclidean body velocity difference $\xi-\mu$ and
    frame-corrected difference $\xi-\Ad_{\delta g}^{-1}(\mu)$.}
    \label{fig:error}
\end{figure}

These group-dependent operators determine how pose and twist uncertainty are coupled.
On $G_{r\ell}$, let $(m,\mu)$ be the group-theoretic mean and $(g,\xi)$ a sample. Equations~\eqref{eq:lsdp} and \eqref{eq:lsdp_inv} give
\begin{align}
    \log\bigl((m,\mu)^{-1}(g,\xi)\bigr)
    =\log\bigl(\delta g,\xi-\Ad_{\delta g}^{-1}(\mu)\bigr),
    \qquad \delta g\defeq m^{-1}g.
    \label{eq:pose_twist_error}
\end{align}
As Fig.~\ref{fig:error} illustrates, $\xi$ and $\mu$ are body twists associated with different poses, namely, $g$ and $m$. 
Subtracting them directly therefore mixes velocity error with frame mismatch. 
The adjoint first expresses the mean twist in the sample body frame, so the subtraction compares twists in the same frame. 
The logarithm then maps this difference to exponential coordinates through $J_r^{-1}(\log\delta g)$, as given by Eq.~\eqref{eq:tse3q_log}.

To complete the description of $G_{r\ell}$, we also derive its closed-form exponential-map Jacobians. Let $x=(\eta,\rho,\omega,\nu)$, $X=[\ad_x]$, and $\theta=\norm{\eta}$. 
Using the coefficients $c_i$ and $\kappa_i$ defined in Appendix~\ref{app:formulas},
\begin{equation}
\begin{aligned}
    J_\ell(x)
    &=I+\left(\frac{c_0-9c_1}{8}+3c_2\right)X
        +\frac{c_1-11c_2+35c_3}{8}X^2
        +\left(\frac{-c_2+7c_3}{4}-3c_4\right)X^3\\
    &\quad+\frac{-c_3+9c_4-21c_5}{4}X^4
        +\left(\frac{c_4-5c_5}{8}+c_6\right)X^5
        +\frac{c_5-7c_6+15c_7}{8}X^6,
    \label{eq:Jl_tse3ql}
\end{aligned}
\end{equation}
\begin{equation}
\begin{aligned}
    J_\ell^{-1}(x)
    &=I-\frac12 X
        +\left(3-\frac{7\kappa_1+15\kappa_2+2\kappa_3}{8}\right)
        \left(\frac{X}{\theta}\right)^2\\
    &\quad+\left(3-\frac{5\kappa_1+5\kappa_2+2\kappa_3}{4}\right)
        \left(\frac{X}{\theta}\right)^4
        +\left(1-\frac{3\kappa_1+3\kappa_2+2\kappa_3}{8}\right)
        \left(\frac{X}{\theta}\right)^6.
    \label{eq:Jl_inv_tse3ql}
\end{aligned}
\end{equation}
The right Jacobians follow by replacing $x$ with $-x$. These formulas are exact finite polynomials in $X$. 
Near $\theta=0$, we use small-angle series only to evaluate their scalar coefficients. 
This avoids truncating the matrix series in $X$ and confines the small-angle approximation to the attitude component $\eta$, rather than the full error vector $x$. 
Appendix~\ref{app:Jl_deriv} gives the derivation.

\subsection{Controlled Rigid-Body Dynamics}

Let $g=(q,\vframe{r}[S])\in\SEq$ denote the pose, $\xi=(\vframe{\Omega}[B],\vframe{v}[B])$ the body twist, and $u=(\vframe{\tau}[B],\vframe{\phi}[B])$ the applied body-frame torque and force. 
Time arguments are suppressed below. For mass $m$ and body inertia $\II$, the forced Euler--Poincar\'{e} equation \cite[Section 1.2]{marsden1998} with an additional pose-dependent body-frame wrench $w(g)$ is
\begin{align}
    \JJ\dot\xi-[\ad_\xi]^\top\JJ\xi
    =u+w(g),
    \qquad
    \JJ\defeq\blkdiag(\II,mI_3).
    \label{eq:ep_eq}
\end{align}

\begin{prop}\label{prop:det_eom}
For $h=(g,\xi)\in G_{r\ell}$ satisfying the reconstruction equation $g^{-1}.\dot g=\xi$ and Eq.~\eqref{eq:ep_eq}, the left-trivialized dynamics are
\begin{align}
    h^{-1}.\dot h
    =\begin{bmatrix}
        \xi\\
        \JJ^{-1}\bigl(u+w(g)+[\ad_\xi]^\top\JJ\xi\bigr)
    \end{bmatrix}.
    \label{eq:eom_general}
\end{align}
\end{prop}
\begin{proof}
By Lemma~\ref{lem:left_tri_tan}, reconstruction gives $(h^{-1}.\dot h)\ve=(\xi,\dot\xi)$ since $\ad_\xi(\xi)=0$. 
Substitution of Eq.~\eqref{eq:ep_eq} gives the result.
\end{proof}

For gravity, $w(g)=(0,m\Ad_{q^*}(\vframe{\gamma}[S]))$, where $\vframe{\gamma}[S]$ is the spatial gravitational acceleration. 
The corresponding left-trivialized drift, evaluated at state $h = (g,\xi)$ and control $u$, is
\begin{align}
    f(h,u,t)\defeq h^{-1}.\dot h
    =\begin{bmatrix}
    \xi\\
    f_\xi(h,u,t)
    \end{bmatrix}
    =\begin{bmatrix}
        \vframe{\Omega}[B]\\
        \vframe{v}[B]\\
        \II^{-1}\bigl(\vframe{\tau}[B]
            -\vframe{\Omega}[B]\ha\II\vframe{\Omega}[B]\bigr)\\
        \dfrac{1}{m}\vframe{\phi}[B]
            +\Ad_{q^*}(\vframe{\gamma}[S])
            -\vframe{\Omega}[B]\ha\vframe{v}[B]
    \end{bmatrix}.
    \label{eq:eom}
\end{align}

\subsection{Stochastic Equations of Motion}

For the stochastic process $h_t=(g_t,\xi_t)$ with control law $u=u(h_t,t)$, write $f^s(h,t)\defeq f(h,u(h,t),t)=(\xi,f_\xi^s(h,t))$.
We assume that noise enters only the twist dynamics, with diffusion $B(g,t)\in\rr^{6\times6}$ independent of $\xi$:
\begin{align}
    \begin{cases}
        \d g_t=(g_t.\xi_t)\,\d t,\\
        \d\xi_t=f_\xi^s(h_t,t)\,\d t+B(g_t,t)\circ\d W_t,
    \end{cases}
    \label{eq:group_sde}
\end{align}
where $W_t$ is a six-dimensional Wiener process and $B$ is smooth in $g$.

\begin{theorem}
\label{thm:sde}\label{prop:sde}\label{prop:equiv_sde}
On either $G_{rd}$ or $G_{r\ell}$, Eq.~\eqref{eq:group_sde} has the left-trivialized Stratonovich representation
\begin{align}
    h_t^{-1}.(\circ\d h_t)
    =f^s(h_t,t)\,\d t+F^s(g_t,t)\circ\d W_t,
    \qquad
    F^s(g,t)=\begin{bmatrix}0_{6\times6}\\B(g,t)\end{bmatrix}.
    \label{eq:lsde_strat}
\end{align}
Its equivalent It\^{o} representation is
\begin{align}
    h_t^{-1}\d h_t
    =f^s(h_t,t)\,\d t+F^s(g_t,t)\,\d W_t.
    \label{eq:lsde_Ito}
\end{align}
\end{theorem}
\begin{proof}
On $G_{r\ell}$, Lemma~\ref{lem:left_tri_tan} and reconstruction give
\begin{align}
    h_t^{-1}.(\circ\d h_t)
    &=\begin{bmatrix}
        \xi_t\,\d t\\
        \circ\d\xi_t+\ad_{\xi_t}(\xi_t)\,\d t
    \end{bmatrix}
    =\begin{bmatrix}\xi_t\,\d t\\\circ\d\xi_t\end{bmatrix}.
\end{align}
On $G_{rd}$, left trivialization acts componentwise and gives the same expression, since left trivialization on the $\rr^6$ component is the identity map. 
Substitution of Eq.~\eqref{eq:group_sde} establishes Eq.~\eqref{eq:lsde_strat} in both cases. For the It\^{o} correction in Eq.~\eqref{eq:Ito_correction}, $F_{ij}^s=0$ for $i=1,\ldots,6$ because no noise enters the pose dynamics. 
For $i=7,\ldots,12$, right multiplication by $\exp(\tau e_i)$ changes only the twist under either product structure. 
Since $F^s(g,t)$ is independent of twist, $E_i^r(F_{kj}^s)=0$ for these indices. Therefore,
\begin{align}
    f-f^s
    =\frac12\left(\sum_{i=1}^{6}E_i^r(F_{kj}^s)F_{ij}^s
    +\sum_{i=7}^{12}E_i^r(F_{kj}^s)F_{ij}^s\right)e_k=0.
\end{align}
Together with $F=F^s$, this gives Eq.~\eqref{eq:lsde_Ito} on both groups.
\end{proof}
Thus, the same physical process has the same left-trivialized drift and diffusion coefficients under both group structures, in either Stratonovich or It\^{o} form.
We henceforth write $f=f^s$ and $F=F^s$, with the control law absorbed into the state and time dependence of $f(h,t)$. Setting $B=0$ also gives the same deterministic dynamics on both groups.

\section{Numerical Experiment}
\sisetup{per-mode=symbol}

\subsection{Lunar Powered Descent}

We use lunar powered descent to assess two effects of group structure: how well concentrated Gaussians parametrized by the group-theoretic mean and covariance describe the state distribution, and how accurately uncertainty is propagated. 
This strongly actuated maneuver is a challenging propagation scenario: large body-frame thrust and torque drive coupled translational and rotational dynamics, while actuation noise continuously injects uncertainty. 
Adapting a scenario from \cite{kamath2023customized}, we generate a nominal trajectory with 15 uniformly spaced nodes over $t\in[0,t_f]$ and $t_f\approx\qty{130}{\second}$. 
A first-order hold on the nodal torque, thrust, and mass defines the prescribed input $u(t)$ and mass parameter $m(t)$ shared by all Monte Carlo (MC) samples and propagation methods. 
All propagation is open-loop, with no measurement updates.

\begin{figure}[htbp]
    \centering
    \includegraphics[width=0.4\linewidth]{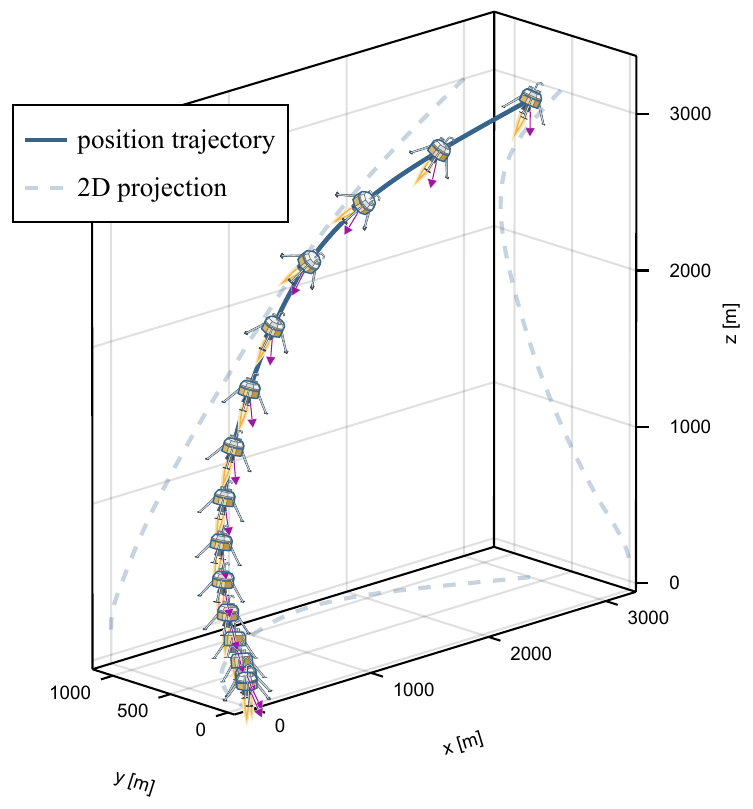}
    \caption{Nominal lunar powered-descent pose trajectory in ENU coordinates.
    The arrows represent the body-fixed LiDAR boresight vector, providing a visual reference for the lander's attitude.}
    \label{fig:powered-descent-pose}
\end{figure}

The spatial frame $\{S\}$ is East-North-Up (ENU), centered at the target landing site. 
Scalar-first quaternions map body coordinates to ENU coordinates, and thrust acts along the positive body $z$-axis. 
Moment of inertia is assumed to be constant. 
Figure~\ref{fig:powered-descent-pose} shows the nominal pose trajectory. 
The scenario, initial uncertainty, and the nominal trajectory constraint parameters can be found at \cite[Table 2 and Appendix B]{scitech2027}.

In addition to $G_{rd}$ and $G_{r\ell}$ from Eq.~\eqref{eq:phase_space_groups}, we consider the direct-product structure
\begin{align}
    G_{dd}\defeq(\SOq\times\rr^3)\times\rr^6
    \label{eq:powered-descent-groups}
\end{align}
implicitly used by the MEKF baseline. Comparing $G_{rd}$ and $G_{r\ell}$ within each Lie-theoretic method isolates the effect of the semidirect pose--twist structure.

Actuation uncertainty enters the twist dynamics through six independent Wiener components. 
Its amplitudes scale with the corresponding prescribed commands, so the diffusion is a function of time only:
\begin{align}
    H(t)\defeq F(g,t)
    =\begin{bmatrix}
        0_{6\times3}&0_{6\times3}\\
        c_\tau\II^{-1}\operatorname{diag}(|\vframe{\tau}[B](t)|)&0_{3\times3}\\
        0_{3\times3}&\dfrac{c_F}{m(t)}
            \operatorname{diag}(|\vframe{\phi}[B](t)|)
    \end{bmatrix},
    \label{eq:powered-descent-noise}
\end{align}
where $c_\tau=c_F=\num{0.05}\sqrt{\unit{\second}}$. Absolute values are componentwise and the corresponding process noise matrix is $Q_c(t)=H(t)H(t)^\top$. 
The same $H(t)$ applies to all three groups because their exponential maps agree for pure twist increments.

We generate $N=1{,}000$ MC trajectories from an initial concentrated Gaussian distribution $h_0\exp(x_0)$ on $G_{rd}$, where $h_0$ is the nominal initial state and $x_0\sim\mathcal N(0,P_0)$, with
\begin{align}
    P_0=\blkdiag \!\left(
        \sigma_\eta^2I_3,\sigma_\rho^2I_3,
        \sigma_\Omega^2I_3,\sigma_v^2I_3\right).
    \label{eq:powered-descent-initial-sampling}
\end{align}
The per-axis initial standard deviations are $\sigma_\eta=\qty{0.5}{\degree}$, $\sigma_\rho=\qty{100}{\meter}$, $\sigma_\Omega=\qty{1e-3}{\radian\per\second}$, and $\sigma_v=\qty{1.5}{\meter\per\second}$.

Theorem~\ref{thm:sde} gives the left-trivialized It\^{o} SDE in Eq.~\eqref{eq:lsde_Ito} on both $G_{rd}$ and $G_{r\ell}$.
We integrate it on $G_{rd}$ using geometric Euler--Maruyama (g-EM) \cite{MarjanovicSolo2018MatrixLieGroupsMadeSimple,MUNIZ2023196} and use the resulting MC ensemble for all group formulations and propagation methods. 
The MC reference uses 25,000 steps, $\Delta t_{\mathrm{MC}}=t_f/25000\approx\qty{0.0052}{\second}$. 
Step refinement with shared Wiener paths gives empirical convergence orders of $1.001$--$1.004$, supporting strong order-$1$ convergence of g-EM for the present model.

\subsection{Uncertainty Representation}
We first assess how well the group-theoretic mean and covariance represent the propagated state distribution by comparing the corresponding concentrated Gaussian approximation with the MC reference.
Using MC-derived moments excludes errors introduced by the moment-propagation algorithms.

\begin{figure}[htbp]
    \centering
    \begin{minipage}[b]{0.494\linewidth}
        \centering
        \raisebox{1ex}{\includegraphics[width=\linewidth,trim=0 50bp 0 35bp,clip]{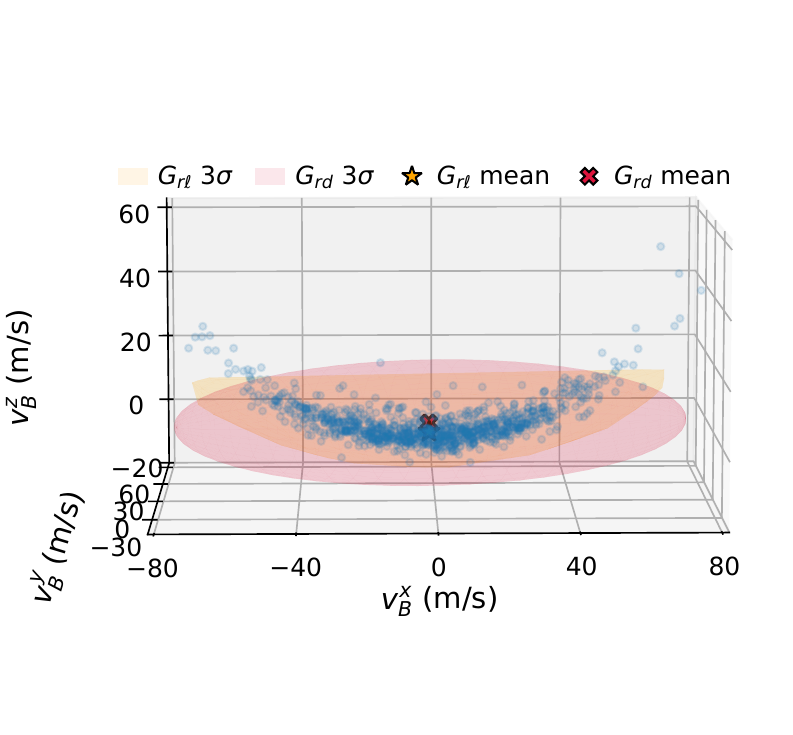}}\\[-0.5ex]
        {\small\textbf{(a)}}
    \end{minipage}\hfill
    \begin{minipage}[b]{0.494\linewidth}
        \centering
        \includegraphics[width=\linewidth]{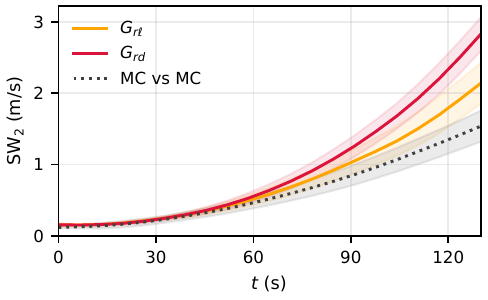}\\[-0.5ex]
        {\small\textbf{(b)}}
    \end{minipage}
    \caption{Body-velocity uncertainty representation: (a) terminal MC
    samples with $3\sigma$ covariance boundaries under $G_{rd}$ and
    $G_{r\ell}$; (b) Average $\mathrm{SW}_2$ between MC and reconstructed body velocity samples. Curves show trial means, with shading illustrating $95\%$ confidence intervals for the means.}
    \label{fig:vB}
\end{figure}

We focus on body velocity, whose banana-shaped distribution provides a clear illustration of the pose--twist coupling addressed by the semidirect-product structure. 
Figure~\ref{fig:vB}a compares its terminal distribution with moment-based representations under $G_{rd}$ and $G_{r\ell}$. 
Since $G_{dd}$ and $G_{rd}$ share an additive $\rr^6$ twist factor, their MC-derived Gaussian approximations of body velocity coincide, so only $G_{rd}$ is shown. Under $G_{rd}$, the velocity log-errors are ordinary body-velocity differences, and the marginal covariance ellipsoid does not capture the banana-shaped distribution. 
Under $G_{r\ell}$, the exponential map bends the pushforward $3\sigma$ boundary to follow the sample spread. 
This is consistent with the relative-pose correction in Eq.~\eqref{eq:pose_twist_error}: accounting for pose--twist coupling allows the mean and covariance to better describe the body-velocity distribution.

Similar behavior has been reported for pose uncertainty: distributions that appear banana-shaped in physical coordinates can be well approximated by Gaussians in exponential coordinates \cite{LongRSS2012Banana,barrau2020imu,vangoor2023eqvio}.

For a quantitative comparison, we sample each group's concentrated Gaussian approximation using its MC-derived group-theoretic moments and extract the body velocities.
We compare these samples with the body velocities of propagated independent MC samples using the empirical sliced 2-Wasserstein distance ($\mathrm{SW}_2$) \cite{nadjahi2020statistical}, a standard metric for comparing two empirical distributions.
We average $\mathrm{SW}_2$ over ten trials with independent concentrated Gaussian draws and MC reference ensembles. 
At each time point, each trial compares $1{,}000$ reconstructed samples with $1{,}000$ propagated independent MC samples using $2^{12}=4{,}096$ shared uniformly sampled projection directions.
The MC--MC comparison averages five comparisons formed from disjoint pairs of the ten independent reference ensembles and provides a baseline for $\mathrm{SW}_2$ arising from finite sampling.
Confidence intervals for the means use Student's $t$ distribution. 
Figure~\ref{fig:vB}b shows that concentrated Gaussian reconstruction on $G_{r\ell}$ more faithfully represents body-velocity uncertainty throughout the descent. 

\subsection{Uncertainty Propagation}

We next assess propagation performance through moment consistency and agreement between concentrated Gaussian approximations based on propagated moments and the MC reference. 
We compare five propagation methods: the MEKF prediction step on $G_{dd}$ \cite{lefferts1982kalman}, an improved UKF-M \cite{Brossard2020UKFM} on $G_{rd}$ and $G_{r\ell}$, and the second-order moment expansion method of \cite{ye2024uncertaintya}, hereafter denoted EM2, on $G_{rd}$ and $G_{r\ell}$.
MEKF uses the Cartesian rigid-body equations with body-frame twists and the standard discrete covariance update via the state-transition matrix (STM)
and state-noise compensation (SNC) \cite{carpenter2018navigation}. Table~\ref{tab:powered-descent-models} summarizes the numerical settings.

We follow the UKF-M prediction procedure \cite{Brossard2020UKFM}, with an improved mean propagation. 
On $G_{rd}$, UKF-M effectively uses the same pose--twist retraction as the $\TSE(3)$ filter in \cite{Mangiacapra2022UKFControlTSE3}: pose errors enter through the $\SEq$ exponential, while body-twist errors are additive. 
On both $G_{rd}$ and $G_{r\ell}$, at each step, we form $2n+1$ state sigma points through the respective exponential map and propagate them using second-order RKMK (RKMK2) on Eq.~\eqref{eq:eom}, with $n=12$. 
The prediction steps in \cite{Brossard2020UKFM,Mangiacapra2022UKFControlTSE3} obtain the predicted mean by propagating the current mean alone. 
We instead compute the weighted group-theoretic mean of all propagated state sigma points using Eq.~\eqref{eq:group_mean} \cite{khan2025means}, allowing the predicted mean to depend on the current uncertainty. 
This modification extends the Euclidean UKF's weighted sigma-point mean update to Lie groups and improves performance in our experiments.
The covariance combines contributions from the propagated state and noise sigma points, expressed as log-errors about this mean.

For EM2, we integrate the mean and covariance equations of \cite{ye2024uncertaintya} together. 
We use RKMK2 for the mean \cite{MuntheKaas1999HighOrderRKManifolds} and a discrete covariance propagation scheme as described in \cite[Appendix C]{scitech2027}.

Each method computes its initial group-theoretic mean and covariance using its assigned group structure and a common set of $100{,}000$ states sampled from the prescribed initial distribution.
\begin{table}[htbp]
\centering
\caption{Propagation settings. All methods use 2500 steps,
$\Delta t=t_f/2500\approx\qty{0.052}{\second}$, and the same prescribed
inputs. UKF-M uses $\alpha=0.5$, $\beta=2$, and $\kappa=0$.}
\label{tab:powered-descent-models}
\small
\renewcommand{\arraystretch}{1.08}
\begin{tabularx}{\linewidth}{@{}llX@{}}
\hline\hline
Method & Group & Integration \\
\midrule
MEKF & $G_{dd}$ & RK4 mean and STM; discrete SNC covariance update \\
UKF-M & $G_{rd}$, $G_{r\ell}$ & RKMK2 sigma-point propagation \\
EM2 & $G_{rd}$, $G_{r\ell}$ & Coupled RKMK2 mean and discrete covariance updates \\
\hline\hline
\end{tabularx}
\end{table}

\begin{figure}[htbp]
    \centering
    \includegraphics[width=\linewidth]{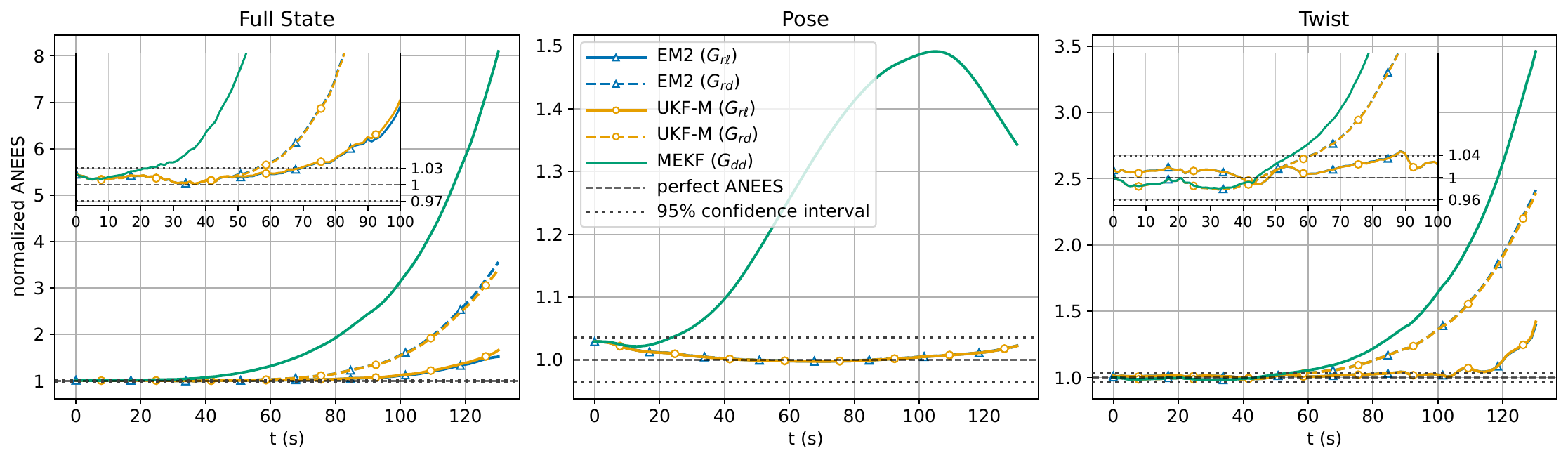}
    \caption{ANEES for the full 12-dimensional state and the
    six-dimensional pose and twist log-error blocks. The standard
    chi-square $95\%$ interval provides an approximate reference
    for MC sampling variability \cite[Sec.~5.4]{BarShalom2001Estimation}.}
    \label{fig:prop_anees}
\end{figure}

We assess moment consistency using dimension-normalized average normalized estimation error squared (ANEES) \cite{brossard2022associating}. For propagated moments $(\mu(t),\Sigma(t))$ and MC samples $h_{(i)}(t)$,
\begin{align}
    \ANEES(t)=\frac{1}{nN}\sum_{i=1}^N
        \log\bigl(\mu(t)^{-1}h_{(i)}(t)\bigr)^\top
        \Sigma(t)^{-1}
        \log\bigl(\mu(t)^{-1}h_{(i)}(t)\bigr).
    \label{eq:anees_prop}
\end{align}
Each method is assessed in its own log-error coordinates. Values near one indicate moment consistency, while values above one indicate overconfidence. Figure~\ref{fig:prop_anees} shows improved consistency with $G_{r\ell}$ for both UKF-M and EM2, most notably in twist and the full state. Pose ANEES remains within the $95\%$ reference interval for all Lie-theoretic methods, which share the same $\SEq$ pose structure. MEKF exhibits the largest departure from one. Despite their distinct propagation mechanisms, UKF-M and EM2 perform similarly on $G_{r\ell}$, and both improve substantially over their direct-product counterparts. Notably, the body twist ANEES of the $G_{r\ell}$ methods remains close to the reference interval until late in the descent.

To assess the combined effect of moment propagation and uncertainty representation, we sample each method's concentrated Gaussian approximation using its propagated group-theoretic mean and covariance. 
We compare these samples with independent propagated MC samples in the common coordinates:
\begin{align}
    \left(
        \log(q_{\mathrm{ref}}(t)^{-1}q),
        \vframe{r}[S],
        \vframe{\Omega}[B],
        \vframe{v}[B]\right)\in\rr^{12},
    \label{eq:powered-descent-wasserstein-metric}
\end{align}
where $q_{\mathrm{ref}}(t)$ is the original MC ensemble's attitude group-theoretic mean on $\SOq$. 
At each time point, we whiten all samples using the original MC ensemble's covariance in these common coordinates to nondimensionalize the errors. 
We then evaluate $\mathrm{SW}_2$ over all 12 coordinates using the same sampling settings and trial-averaging procedure described previously.

\begin{figure}[htbp]
    \centering
    \includegraphics[width=0.5\linewidth]{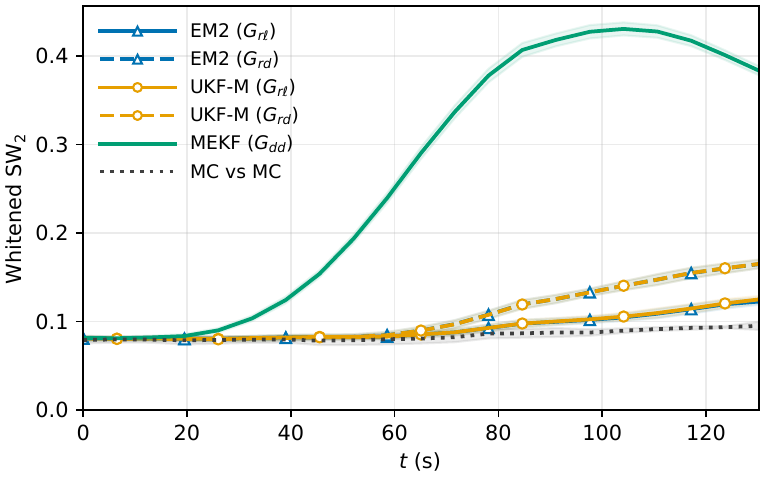}
    \caption{Full-state whitened $\mathrm{SW}_2$ between concentrated Gaussian samples generated from propagated moments and independent MC samples. Averaging and $95\%$ confidence intervals follow Fig.~\ref{fig:vB}b.}
    \label{fig:prop_wasserstein}
\end{figure}

Figure~\ref{fig:prop_wasserstein} shows that, over the latter half of the trajectory, both UKF-M and EM2 reproduce the MC distribution more closely on $G_{r\ell}$ than on $G_{rd}$. The two methods perform nearly identically on $G_{r\ell}$, with their difference much smaller than the improvement from changing group structure. Together, the ANEES and $\mathrm{SW}_2$ results indicate that group structure is the dominant factor for propagation performance in this experiment. The MC-derived representations above show why this choice also matters independently of propagation: the semidirect coordinates allow the MC-derived mean and covariance to capture the body-velocity distribution more faithfully.

Finally, we compare the computational cost of EM2 and UKF-M on $G_{r\ell}$ with that of the MEKF baseline.
All timings were obtained on an Intel Core Ultra 7 165H processor with 32 GB RAM using double-precision JAX with just-in-time (JIT) compilation. 
Compilation and a full warm-up trial were performed prior to timing. 
Each method was evaluated over 2,500 propagation steps. Every step was timed individually and synchronized before being passed to the next step. 
We report the mean wall-clock time per step and the standard deviation (SD) across all 2,500 steps in Table~\ref{tab:powered_descent_runtime}.
\begin{table}[h]
    \centering
    \caption{Runtime per propagation step for one warmed trial.}
    \label{tab:powered_descent_runtime}
    \begin{tabular}{@{}lrr@{}}
        \hline\hline
        Method & Mean ($\mu$s) & SD ($\mu$s) \\
        \hline
        EM2 ($G_{r\ell}$)   & 221.864 & 30.577 \\
        UKF-M ($G_{r\ell}$) &  89.692 & 29.904 \\
        MEKF               &  37.760 & 18.301 \\
        \hline\hline
    \end{tabular}
\end{table}
For this experiment and the present implementations, UKF-M on $G_{r\ell}$ provides a favorable balance between computational cost and uncertainty-propagation performance. 
Its mean per-step runtime is roughly $2.4$ times that of MEKF, while maintaining good agreement with the Monte Carlo reference. 
EM2 provides an accurate, tuning-free alternative based on analytic moment equations, at a higher computational cost.
MEKF achieves the lowest runtime, but this computational advantage comes at a significant accuracy tradeoff.
\section{Conclusion}

Equivalent formulations of stochastic rigid-body dynamics need not yield equally accurate concentrated Gaussian approximations of the propagated distribution. 
For the noise model considered here, the direct- and semidirect-product formulations describe the same physical stochastic process, but their group-theoretic means and covariances parameterize different concentrated Gaussian approximations. 
The choice of group structure is therefore part of the uncertainty approximation itself.

Left trivialization aligns the formulation with the body-frame actuation dynamics and noise.
In the strongly actuated lunar powered-descent experiment, the proposed group structure improves the concentrated Gaussian representation of body-velocity uncertainty and the propagation of full-state uncertainty. 
Its frame-correcting exponential coordinates allow the group-theoretic mean and covariance to better represent the banana-shaped body-velocity distribution. 
Despite their highly distinct propagation mechanisms, UKF-M and EM2 achieve nearly the same performance on the semidirect group. 
The performance of uncertainty propagation in the experiment is largely determined by the group structure instead.

These results support seeking group structures in which Gaussian approximations in log-error coordinates remain accurate as a path toward distributional control.
Accurate propagation of low-order moments is more useful for uncertainty propagation when those moments also provide an adequate representation of the underlying distribution. 
A useful formulation must therefore both propagate the moments accurately and preserve their ability to approximate the distribution. 
Future work will incorporate the proposed structure into stochastic 6DOF trajectory optimization and feedback control, examining whether these advantages improve control of the state distribution.

\appendix
\crefalias{section}{appendix}
\section{Useful Formulas for $\SO(3)$ and $\SE(3)$} \label{app:formulas}
We use the Rodrigues formulas for $\SO(3)$ and $\SE(3)$ from \cite{muller2021review}. Let $\eta,\rho\in\rr^3$ and $\theta\defeq\norm{\eta}$. Write $c_n=c_n(\theta^2)$ for the Stumpff functions,
\begin{align}
    c_n(\theta^2)=\sum_{k=0}^{\infty}\frac{(-1)^k\theta^{2k}}{(n+2k)!},
    \qquad
    c_{n+2}=\frac{1/n!-c_n}{\theta^2},\quad n\ge0.
    \label{eq:stumpff}
\end{align}
In particular, $c_0=\cos\theta$, $c_1=\sinc\theta$, and $c_2=\frac12\sinc^2(\theta/2)$, where $\sinc\theta=\sin\theta/\theta$. We also use $\kappa_1\defeq1/(2c_2)$, $\kappa_2\defeq c_1/(2c_2)$, and $\kappa_3\defeq\kappa_1\kappa_2$. At $\theta=0$, we use the continuous limits of the scalar coefficients. Near zero, we evaluate them using small-angle series to avoid cancellation.

For $\SO(3)$,
\begin{align}
    \exp(\eta)&=I+c_1\eta\ha+c_2(\eta\ha)^2,
    \label{eq:so3_exp}\\
    J_\ell(\eta)&=I+c_2\eta\ha+c_3(\eta\ha)^2,\qquad
    J_\ell^{-1}(\eta)=I-\frac12\eta\ha
        +\frac{1-\kappa_2}{\theta^2}(\eta\ha)^2,\\
    \log(R)&=\left(\frac{R-R^\top}{2c_1}\right)\ve,
        \qquad\theta=\arccos\left(\frac{\tr R-1}{2}\right),
    \label{eq:so3_log}\\
    J_r(\eta)&=J_\ell(-\eta)=J_\ell(\eta)^\top,\qquad
    J_r^{-1}(\eta)=J_\ell^{-1}(-\eta)=J_\ell^{-1}(\eta)^\top.
\end{align}
The exponential maps and Jacobians are defined for all $\eta$. The inverse Jacobians exist when $\theta\ne2k\pi$, $k=1,2,\ldots$. We use the principal rotation-vector logarithm, with $0\le\theta<\pi$. These domains also apply to the following $\SE(3)$ formulas.

For $(R,r)\in\SE(3)$,
\begin{align}
    &[\Ad_{(R,r)}]
    =\begin{bmatrix}R&0\\r\ha R&R\end{bmatrix},
    \qquad \qquad \quad
    [\ad_{(\eta,\rho)}]
    =\begin{bmatrix}\eta\ha&0\\\rho\ha&\eta\ha\end{bmatrix},\\
    &\exp(\eta,\rho)=
        \begin{bmatrix}
        \exp(\eta) & J_\ell(\eta)\rho\\
        0 &1
        \end{bmatrix},
        \qquad 
    \log(R,r)=
        \begin{bmatrix}
        \log(R)\\J_\ell^{-1}(\log(R))r
        \end{bmatrix}.
\end{align}
The lower-left block of the left Jacobian is
\begin{align}
    Q(\eta,\rho)
    =c_2\rho\ha+c_3(\eta\ha\rho\ha+\rho\ha\eta\ha)
        +(\eta\cdot\rho)\bigl((2c_4-c_3)\eta\ha
        +(3c_5-c_4)(\eta\ha)^2\bigr).
\end{align}
Then
\begin{align}
    J_\ell(\eta,\rho)&=
        \begin{bmatrix}J_\ell(\eta)&0\\Q(\eta,\rho)&J_\ell(\eta)\end{bmatrix},
    \qquad \qquad 
    J_\ell^{-1}(\eta,\rho)=
        \begin{bmatrix}J_\ell^{-1}(\eta)&0\\
        -J_\ell^{-1}(\eta)Q(\eta,\rho)J_\ell^{-1}(\eta)&J_\ell^{-1}(\eta)
        \end{bmatrix},
\end{align}
where
\begin{align}
    -J_\ell^{-1}(\eta)Q(\eta,\rho)J_\ell^{-1}(\eta)
    =-\frac12\rho\ha
        +\frac{1-\kappa_2}{\theta^2}(\eta\ha\rho\ha+\rho\ha\eta\ha)
        +\frac{\eta\cdot\rho}{\theta^4}
        (\kappa_1+\kappa_2-2)(\eta\ha)^2.
\end{align}
The inverse right Jacobian is $J_r^{-1}(\eta,\rho)=J_\ell^{-1}(-\eta,-\rho)$, and $Q(-\eta,-\rho)=Q(\eta,\rho)^\top$.

\section{Derivation of Eqs.~\eqref{eq:Jl_tse3ql} and \eqref{eq:Jl_inv_tse3ql}}
\label{app:Jl_deriv}
We derive the Rodrigues formulas by expressing each Jacobian as a polynomial in the adjoint matrix. The \defemph{minimal polynomial} of $X$ is the monic polynomial $m(z)$ of lowest degree $d$ such that $m(X)=0$. An analytic matrix function $f(X)$ can then be written as $r(X)$, where $r(z)$ has degree at most $d-1$ \cite{dummit2004abstract,HighamLin2015MatrixFunctionsShortCourse}. To determine its coefficients, we use Hermite interpolation: at each root of $m(z)$ with multiplicity $a$, we match the value of $f$ and its first $a-1$ derivatives. These $d$ conditions uniquely determine $r(z)$ and ensure that $r(X)=f(X)$.

For $x=(\eta,\rho,\omega,\nu)$, let $X=[\ad_x]$ and $\theta=\norm{\eta}$. To identify the generic minimal polynomial, we evaluate $X$ at randomly chosen components of $x$ and compute the resulting matrix's minimal polynomial in \texttt{SymPy}. This suggests the general form
\begin{align}
    m(z)=z(z^2+\theta^2)^3.
\end{align}
Substitution into Eq.~\eqref{eq:tse3q_ad} verifies $m(X)=0$ for arbitrary $x$. Since $m(z)$ has degree seven, each Jacobian can be expressed as a polynomial in $X$ of degree at most six. For $\theta>0$, $m(z)$ has a simple root at zero and roots $\pm i\theta$ of multiplicity three. The seven conditions are
\begin{align}
    r(0)=f(0),\qquad
    r^{(k)}(\pm i\theta)=f^{(k)}(\pm i\theta),\quad k=0,1,2.
\end{align}
For $J_\ell$, Eq.~\eqref{eq:mat_Jl} gives $f(z)=(e^z-1)/z$, with $f(0)=1$. Solving these conditions in \texttt{SymPy} and writing the coefficients in terms of the Stumpff functions gives Eq.~\eqref{eq:Jl_tse3ql}. Repeating the calculation with $f(z)=z/(e^z-1)$ gives Eq.~\eqref{eq:Jl_inv_tse3ql}, provided $\theta\ne2k\pi$, $k=1,2,\ldots$. When $\theta=0$, the same block structure gives $X^3=0$. The Jacobian series in Eqs.~\eqref{eq:mat_Jl} and \eqref{eq:mat_Jl_inv} therefore terminate after $X^2$.
\section*{Acknowledgments}

This work was supported by NASA Space Technology Graduate Research Opportunity 80NSSC25K0234. The authors thank David Woffinden and James McCabe for ideas on uncertainty propagation applications and best practices, John Etnyre for mathematical discussions, and Jikai Ye for help with the EM2 algorithm. OpenAI ChatGPT was used to assist with improving the flow, readability, and visualization of author-written text, and with the development of numerical code. The authors reviewed all AI-assisted outputs and take full responsibility for the final results.


\bibliography{references}

@ARTICLE{ye2024uncertaintya,
  author={Ye, Jikai and Jayaraman, Amitesh S. and Chirikjian, Gregory S.},
  journal={IEEE Transactions on Automatic Control}, 
  title={Uncertainty Propagation on Unimodular Matrix Lie Groups}, 
  year={2025},
  volume={70},
  number={4},
  pages={2450-2465},
  doi={10.1109/TAC.2024.3486652}}

@INPROCEEDINGS{ye2024uncertaintyb,
  author={Ye, Jikai and Chirikjian, Gregory S.},
  booktitle={2024 IEEE 63rd Conference on Decision and Control (CDC)}, 
  title={Uncertainty Propagation and Bayesian Fusion on Unimodular Lie Groups from a Parametric Perspective}, 
  year={2024},
  volume={},
  number={},
  pages={8702-8709},
  doi={10.1109/CDC56724.2024.10886748}}

@book{chirikjian2012stochastic,
  author    = {Chirikjian, Gregory S.},
  title     = {Stochastic Models, Information Theory, and Lie Groups, Volume 2: Analytic Methods and Modern Applications},
  series    = {Applied and Numerical Harmonic Analysis},
  publisher = {Birkh{\"a}user Boston},
  address   = {Boston, MA},
  year      = {2012},
  doi       = {10.1007/978-0-8176-4944-9},
  isbn      = {978-0-8176-4943-2}
}

@book{marsden1999introduction,
  title={Introduction to Mechanics and Symmetry: a Basic Exposition of Classical Mechanical Systems},
  author={Marsden, Jerrold E and Ratiu, Tudor S},
  year={1999},
  publisher={Springer New York, NY},
  edition={Second},
  doi={10.1007/978-0-387-21792-5}
}

@book{marsden1998,
    author = {Marsden,Jerrold E. and Ratiu, Tudor S.},
    title = {Mechanics and Symmetry: Reduction Theory},
    publisher = {Online},
    year = {1998},
    url = {http://inis.jinr.ru/sl/P_Physics/PC_Classical%20physics/PCft_Classical%20fields/Marsden%20J.%20Mechanics%20and%20symmetry.%20Reduction%20theory%20(book%202,%20draft%201998)(145s).pdf}
}

@book{dummit2004abstract,
  title={Abstract Algebra},
  author={Dummit, David Steven and Foote, Richard M},
  edition={3},
  year={2004},
  publisher={Wiley Hoboken, NJ}
}

@book{kirillov2008introduction,
  title={An Introduction to Lie Groups and Lie Algebras},
  author={Kirillov, Jr., Alexander A},
  volume={113},
  series={Cambridge Studies in Advanced Mathematics},
  year={2017},
  publisher={Cambridge University Press},
  doi={10.1017/CBO9780511755156}
}

@book{holm2011geometric,
  title={Geometric Mechanics-Part II: Rotating, Translating and Rolling},
  author={Holm, Darryl D},
  year={2011},
  publisher={Imperial College Press},
  edition={2},
  doi={10.1142/p802}
}

@article{muller2021review,
    author = {Müller, Andreas},
    title = {Review of the exponential and Cayley map on SE(3) as relevant for Lie group integration of the generalized Poisson equation and flexible multibody systems},
    journal = {Proceedings of the Royal Society A: Mathematical, Physical and Engineering Sciences},
    volume = {477},
    number = {2253},
    pages = {20210303},
    year = {2021},
    month = {09},
    issn = {1364-5021},
    doi = {10.1098/rspa.2021.0303},
    eprint = {https://royalsocietypublishing.org/rspa/article-pdf/doi/10.1098/rspa.2021.0303/726267/rspa.2021.0303.pdf},
}

@article{moskowitz2003exponential,
     author = {Martin Moskowitz and Richard Sacksteder},
     title = {The {Exponential} {Map} and {Differential} {Equations} on {Real} {Lie} {Groups}},
     journal = {Journal of Lie Theory},
     pages = {291--306},
     year = {2003},
     volume = {13},
     number = {1},
     doi = {10.5802/jolt.304},
     zbl = {1017.22005},
}

@article{markley2004attitude,
  title={Attitude estimation or quaternion estimation?},
  author={Markley, F Landis},
  journal={The Journal of the Astronautical Sciences},
  volume={52},
  number={1-2},
  pages={221--238},
  year={2004},
  publisher={Springer},
  DOI={10.1007/BF03546430}
}

@article{MuntheKaas1999HighOrderRKManifolds,
  author  = {Hans Munthe-Kaas},
  title   = {High order Runge-Kutta methods on manifolds},
  journal = {Applied Numerical Mathematics},
  volume  = {29},
  number  = {1},
  pages   = {115--127},
  year    = {1999},
  month   = jan,
  doi     = {10.1016/S0168-9274(98)00030-0}
}

@misc{khan2025means,
  author       = {Khan, Shiraz and Ye, Jikai and Chirikjian, Gregory S.},
  title        = {Means of Random Variables in {Lie} Groups},
  howpublished = {arXiv:2508.12030},
  month        = aug,
  year         = {2025},
  doi          = {10.48550/arXiv.2508.12030}
}

@book{hsu2002stochastic,
  author    = {Hsu, Elton P.},
  title     = {Stochastic Analysis on Manifolds},
  series    = {Graduate Studies in Mathematics},
  volume    = {38},
  publisher = {American Mathematical Society},
  address   = {Providence, RI},
  year      = {2002},
  isbn      = {978-0-8218-0802-3},
  doi       = {10.1090/gsm/038}
}

@techreport{CatuognoRuffino2003StochasticExponential,
  author      = {Catuogno, Pedro J. and Ruffino, Paulo R. C.},
  title       = {Stochastic Exponential in {Lie} Groups and Its Applications},
  institution = {Institute of Mathematics, Statistics and Scientific Computing, Universidade Estadual de Campinas},
  type        = {Research Report},
  number      = {7/2003},
  address     = {Campinas, Brazil},
  year        = {2003},
  url         = {https://www.ime.unicamp.br/sites/default/files/pesquisa/relatorios/rp-2003-7.pdf}
}

@article{Mangiacapra2022UKFControlTSE3,
  author  = {Gennaro Mangiacapra and Matthew Wittal and Elisa Capello and Morad Nazari},
  title   = {Unscented Kalman Filter and Control on {TSE}(3) with Application to Spacecraft Dynamics},
  journal = {Nonlinear Dynamics},
  year    = {2022},
  volume  = {108},
  number  = {3},
  pages   = {2127--2146},
  doi     = {10.1007/s11071-022-07293-x}
}

@inproceedings{Fornasier2022EqFInputBiases,
  author    = {Alessandro Fornasier and Yonhon Ng and Robert Mahony and Stephan Weiss},
  title     = {Equivariant Filter Design for Inertial Navigation Systems with Input Measurement Biases},
  booktitle = {Proceedings of the 2022 IEEE International Conference on Robotics and Automation (ICRA)},
  year      = {2022},
  pages     = {4333--4339},
  doi       = {10.1109/ICRA46639.2022.9811778}
}

@article{Fornasier2025EquivariantSymmetriesINS,
  author  = {Alessandro Fornasier and Yixiao Ge and Pieter van Goor and Robert Mahony and Stephan Weiss},
  title   = {Equivariant Symmetries for Inertial Navigation Systems},
  journal = {Automatica},
  year    = {2025},
  volume  = {181},
  pages   = {112495},
  doi     = {10.1016/j.automatica.2025.112495}
}

@inproceedings{Jayaraman2023LieTheoreticUncertainty,
  author    = {Amitesh S. Jayaraman and Jikai Ye and Gregory S. Chirikjian},
  title     = {A Lie-Theoretic Approach to Propagating Uncertainty Jointly in Attitude and Angular Momentum},
  booktitle = {Proceedings of the 62nd IEEE Conference on Decision and Control (CDC)},
  year      = {2023},
  pages     = {3212--3219},
  doi       = {10.1109/CDC49753.2023.10383678}
}

@incollection{PennecArsigny2013ExponentialBarycenters,
  author    = {Xavier Pennec and Vincent Arsigny},
  title     = {Exponential Barycenters of the Canonical Cartan Connection and Invariant Means on Lie Groups},
  booktitle = {Matrix Information Geometry},
  editor    = {Frank Nielsen and Rajendra Bhatia},
  pages     = {123--166},
  year      = {2013},
  publisher = {Springer},
  address   = {Berlin, Heidelberg},
  doi       = {10.1007/978-3-642-30232-9_7}
}

@article{MarjanovicSolo2018MatrixLieGroupsMadeSimple,
  author  = {Goran Marjanovic and Victor Solo},
  title   = {Numerical Methods for Stochastic Differential Equations in Matrix Lie Groups Made Simple},
  journal = {IEEE Transactions on Automatic Control},
  volume  = {63},
  number  = {12},
  pages   = {4035--4050},
  year    = {2018},
  doi     = {10.1109/TAC.2018.2798703}
}

@INPROCEEDINGS{Brossard2020UKFM,
  author={Brossard, Martin and Barrau, Axel and Bonnabel, Silvère},
  booktitle={2020 IEEE International Conference on Robotics and Automation (ICRA)}, 
  title={A Code for Unscented Kalman Filtering on Manifolds (UKF-M)}, 
  year={2020},
  volume={},
  number={},
  pages={5701-5708},
  doi={10.1109/ICRA40945.2020.9197489}}

@article{Norris1992,
  author    = {Norris, James R.},
  title     = {A complete differential formalism for stochastic calculus in manifolds},
  journal   = {S{\'e}minaire de probabilit{\'e}s de Strasbourg},
  volume    = {26},
  year      = {1992},
  pages     = {189--209},
  publisher = {Springer},
  doi       = {10.1007/BFb0084322}
}

@inproceedings{LongRSS2012Banana,
  author    = {Andrew Long and Kevin Wolfe and Michael Mashner and Gregory Chirikjian},
  title     = {The Banana Distribution is Gaussian: A Localization Study with Exponential Coordinates},
  booktitle = {Proceedings of Robotics: Science and Systems VIII},
  year      = {2012},
  address   = {Sydney, Australia},
  month     = jul,
  doi       = {10.15607/RSS.2012.VIII.034}
}

@article{brossard2022associating,
  author  = {Martin Brossard and Axel Barrau and Paul Chauchat and Silv{\`e}re Bonnabel},
  title   = {Associating Uncertainty to Extended Poses for on Lie Group IMU Preintegration With Rotating Earth},
  journal = {IEEE Transactions on Robotics},
  volume  = {38},
  number  = {2},
  pages   = {998--1015},
  year    = {2022},
  month   = apr,
  doi     = {10.1109/TRO.2021.3100156}
}

@article{Bourmaud2015CDLGEKF,
  author  = {Guillaume Bourmaud and R{\'e}mi M{\'e}gret and Marc Arnaudon and Audrey Giremus},
  title   = {Continuous-Discrete Extended Kalman Filter on Matrix Lie Groups Using Concentrated Gaussian Distributions},
  journal = {Journal of Mathematical Imaging and Vision},
  volume  = {51},
  number  = {1},
  pages   = {209--228},
  year    = {2015},
  month   = jan,
  doi     = {10.1007/s10851-014-0517-0}
}

@INPROCEEDINGS{barrau2020imu,
  author={Barrau, Axel and Bonnabel, Silvère},
  booktitle={2020 IEEE International Conference on Robotics and Automation (ICRA)}, 
  title={A Mathematical Framework for IMU Error Propagation with Applications to Preintegration}, 
  year={2020},
  volume={},
  number={},
  pages={5732-5738},
  doi={10.1109/ICRA40945.2020.9197492}}

@article{vangoor2023eqvio,
  author    = {Pieter van Goor and Robert Mahony},
  title     = {EqVIO: An Equivariant Filter for Visual-Inertial Odometry},
  journal   = {IEEE Transactions on Robotics},
  volume    = {39},
  number    = {5},
  pages     = {3567--3585},
  year      = {2023},
  doi       = {10.1109/TRO.2023.3289587}
}

@article{lefferts1982kalman,
  author  = {E. J. Lefferts and F. Landis Markley and M. D. Shuster},
  title   = {Kalman Filtering for Spacecraft Attitude Estimation},
  journal = {Journal of Guidance, Control, and Dynamics},
  volume  = {5},
  number  = {5},
  pages   = {417--429},
  year    = {1982},
  doi     = {10.2514/3.56190}
}

@incollection{HighamLin2015MatrixFunctionsShortCourse,
  author    = {Higham, Nicholas J. and Lin, Lijing},
  title     = {Matrix Functions: A Short Course},
  booktitle = {Matrix Functions and Matrix Equations},
  editor    = {Bai, Zhaojun and Gao, Weiguo and Su, Yangfeng},
  series    = {Series in Contemporary Applied Mathematics},
  volume    = {19},
  publisher = {World Scientific},
  address   = {Singapore},
  year      = {2015},
  pages     = {1--27},
  doi       = {10.1142/9789814675772_0001}
}

@article{MUNIZ2023196,
title = {Strong stochastic Runge-Kutta–Munthe-Kaas methods for nonlinear Itô SDEs on manifolds},
journal = {Applied Numerical Mathematics},
volume = {193},
pages = {196-203},
year = {2023},
issn = {0168-9274},
doi = {10.1016/j.apnum.2023.07.024},
author = {Michelle Muniz and Matthias Ehrhardt and Michael Günther and Renate Winkler}
}

@INPROCEEDINGS{Ridderhof2019,
  author={Ridderhof, Jack and Okamoto, Kazuhide and Tsiotras, Panagiotis},
  booktitle={2019 IEEE 58th Conference on Decision and Control (CDC)}, 
  title={Nonlinear Uncertainty Control with Iterative Covariance Steering}, 
  year={2019},
  volume={},
  number={},
  pages={3484-3490},
  doi={10.1109/CDC40024.2019.9029993}}

@techreport{carpenter2018navigation,
  author      = {Carpenter, J. Russell and D'Souza, Christopher N.},
  title       = {Navigation Filter Best Practices},
  institution = {NASA Engineering and Safety Center},
  address     = {Hampton, VA},
  type        = {Technical Publication},
  number      = {NASA/TP-2018-219822},
  year        = {2018},
  month       = apr,
  url         = {https://ntrs.nasa.gov/citations/20180003657}
}

@book{BarShalom2001Estimation,
  author    = {Bar-Shalom, Yaakov and Li, X. Rong and Kirubarajan, Thiagalingam},
  title     = {Estimation with Applications to Tracking and Navigation: Theory, Algorithms and Software},
  publisher = {John Wiley \& Sons},
  year      = {2001},
  doi       = {10.1002/0471221279}
}

@inproceedings{nadjahi2020statistical,
  author    = {Nadjahi, Kimia and Durmus, Alain and Chizat, L{\'e}na{\"i}c and Kolouri, Soheil and Shahrampour, Shahin and {\c S}im{\c s}ekli, Umut},
  title     = {Statistical and Topological Properties of Sliced Probability Divergences},
  booktitle = {Advances in Neural Information Processing Systems},
  volume    = {33},
  pages     = {20802-20812},
  year      = {2020},
  url       = {https://proceedings.neurips.cc/paper_files/paper/2020/hash/eefc9e10ebdc4a2333b42b2dbb8f27b6-Abstract.html}
}

@incollection{yu2025covariance,
  author    = {Hongzhe Yu and Zhenyang Chen and Yongxin Chen},
  title     = {Covariance Steering for Nonlinear Control-Affine Systems},
  booktitle = {Geometry, Topology and Control System Design:
               Proceedings of a Banff International Research Station Workshop},
  editor    = {Mohamed Ali Belabbas},
  series    = {AIMS on Applied Mathematics},
  volume    = {13},
  pages     = {181--198},
  publisher = {American Institute of Mathematical Sciences},
  year      = {2025},
  isbn      = {978-1-60133-027-7}
}

@inproceedings{ridderhof2018uncertainty,
author = {Jack Ridderhof and Panagiotis Tsiotras},
title = {Uncertainty Quantification and Control During Mars Powered Descent and Landing using Covariance Steering},
booktitle = {AIAA Guidance, Navigation, and Control Conference},
year = 2018,
doi = {10.2514/6.2018-0611},
eprint = {https://arc.aiaa.org/doi/pdf/10.2514/6.2018-0611}
}

@article{Zhang2023Stochastic6DOFRendezvous,
  author  = {Yanquan Zhang and Min Cheng and Bin Nan and Shunli Li},
  title   = {Stochastic Trajectory Optimization for 6-DOF Spacecraft Autonomous Rendezvous and Docking with Nonlinear Chance Constraints},
  journal = {Acta Astronautica},
  volume  = {208},
  pages   = {62--73},
  year    = {2023},
  doi     = {10.1016/j.actaastro.2023.04.004}
}

@article{Cesic2017ExtendedInformationFilter,
  author  = {Josip {\'C}esi{\'c} and Ivan Markovi{\'c} and Mario Bukal and Ivan Petrovi{\'c}},
  title   = {Extended Information Filter on Matrix Lie Groups},
  journal = {Automatica},
  volume  = {82},
  pages   = {226--234},
  year    = {2017},
  doi     = {10.1016/j.automatica.2017.04.056}
}

@article{Iserles2000LieGroupMethods,
  author  = {Arieh Iserles and Hans Z. Munthe-Kaas and Syvert P. N{\o}rsett and Antonella Zanna},
  title   = {Lie-Group Methods},
  journal = {Acta Numerica},
  volume  = {9},
  pages   = {215--365},
  year    = {2000},
  doi     = {10.1017/S0962492900002154}
}

@inproceedings{kamath2023customized,
  author    = {Kamath, Abhinav G. and Elango, Purnanand and Kim, Taewan and McEowen, Skye and Mesbahi, Mehran and A{\c{c}}{\i}kme{\c{s}}e, Beh{\c{c}}et and Yu, Yue and Carson, John M.},
  title     = {Customized Real-Time First-Order Methods for Onboard Dual Quaternion-based 6-DoF Powered-Descent Guidance},
  booktitle = {AIAA SciTech 2023 Forum},
  year      = {2023},
  publisher = {American Institute of Aeronautics and Astronautics},
  doi       = {10.2514/6.2023-2003}
}

@article{dantam2021,
author = {Neil T Dantam},
title ={Robust and efficient forward, differential, and inverse kinematics using dual quaternions},

journal = {The International Journal of Robotics Research},
volume = {40},
number = {10-11},
pages = {1087-1105},
year = {2021},
doi = {10.1177/0278364920931948},
eprint = { 
        https://doi.org/10.1177/0278364920931948
}
}

@inproceedings{scitech2027,
  author    = {Zhuochen Wang and Panagiotis Tsiotras},
  title     = {On Uncertainty Propagation for {6DOF} Rigid-Body Motion},
  booktitle = {AIAA SciTech 2027 Forum},
  year      = {2027},
  month     = jan,
  publisher = {American Institute of Aeronautics and Astronautics},
  note      = {To appear}
}


\end{document}